\documentclass[11pt]{article}
\pdfoutput=1
\usepackage[dvips]{graphicx}
\usepackage{epsfig,amsmath,amssymb,amsbsy,amstext,amsfonts,verbatim,mathrsfs,array,layout,textcomp,latexsym,slashed,graphicx,bbm,physics,tensor,cite}
\usepackage[english]{babel}
\usepackage{bm}
\usepackage{xcolor}
\usepackage[normalem]{ulem}
\usepackage{appendix}
\usepackage{enumitem}
\usepackage{rotating}
\usepackage[hidelinks]{hyperref}
\usepackage{mathrsfs}

\usepackage{soul}
\usepackage{dsfont}
\usepackage{tikz-feynman}
\tikzset{
	cross/.style={path picture={\draw[black]
			(path picture bounding box.south east) -- (path picture bounding box.north west)
			(path picture bounding box.south west) -- (path picture bounding box.north east);}}
}

\allowdisplaybreaks

\makeatletter

\newenvironment{bottompar}{\par\vspace*{\fill}}{\clearpage}
\newcommand*{\xdash}[1][3em]{\rule[0.5ex]{#1}{0.55pt}}
\def\mysection#1{{\bf #1.} }

\def\mysection#1{{\bf #1.}}

\definecolor{awesome}{rgb}{1.0, 0.13, 0.32}

\definecolor{airforceblue}{rgb}{0.36, 0.48, 0.84}
\definecolor{purplemath}{rgb}{0.5, 0, 0.5}

\newcommand{\be}{\begin{equation}}
\newcommand{\ee}{\end{equation}}
\newcommand{\bea}{\begin{eqnarray}}
\newcommand{\eea}{\end{eqnarray}}

\newcommand*\diff{\mathop{}\!\mathrm{d}}

\newcommand{\eq}[1]{(\ref{#1})}

\begin{document}

\begin{titlepage}
	\setcounter{page}{1} \baselineskip=15.5pt 
	\thispagestyle{empty}
	$\quad$
	{\raggedleft IFT UAM-CSIC 24-183\\ DESY-24-205 \par}
	\vskip 60 pt
	
	\begin{center}
		{\fontsize{17.28}{17.28} \bf Intrinsic non-Gaussianity of ultra slow-roll inflation}
	\end{center}

	\vskip 20pt
	\begin{center}
		\noindent
		{\fontsize{12}{30}\selectfont  Guillermo Ballesteros$^{1,2}$, 
		Jes\'us Gamb\'in Egea$^{2}$, 
		Thomas Konstandin$^{3}$,\\
Alejandro P\'erez Rodr\'iguez$^{1,2}$, 
Mathias Pierre$^{3}$,  and
Juli\'an Rey$^{3}$}
	\end{center}

	\begin{center}
		\vskip 4pt
		\textit{ $^1${Departamento de F\'{\i}sica Te\'{o}rica, Universidad Aut\'{o}noma de Madrid (UAM), \\Campus de Cantoblanco, 28049 Madrid, Spain}
		}
		\vskip 5pt
		\textit{ $^2${Instituto de F\'{\i}sica Te\'{o}rica UAM-CSIC,  Campus de Cantoblanco, 28049 Madrid, Spain}
		\vskip 5pt
		\textit{  $^3$Deutsches Elektronen-Synchrotron DESY, Notkestr. 85, 22607 Hamburg, Germany}
		}
	\end{center}
	
	\vspace{0.4cm}
	\centerline{\bf Abstract}
	\vspace{0.3cm}
	\noindent 
	
	We study the non-Gaussian tail of the curvature fluctuation, $\zeta$, in an inflationary scenario with a transient ultra slow-roll phase that generates a localized large enhancement of the spectrum of $\zeta$. To do so, we implement a numerical procedure that provides the probability distribution of $\zeta$ order by order in perturbation theory. The non-Gaussianities of $\zeta$ can be shown to arise from its non-linear relation to the inflaton fluctuations and from the {\it intrinsic} non-Gaussianities of the latter, which stem from its self interactions. We find that intrinsic non-Gaussianities, which have often been ignored to estimate the abundance of primordial black holes in this kind of scenario, are important. The relevance of the intrinsic contribution depends on the rapidity with which the transient ultra slow-roll phase occurs, as well as on its duration. Our method cannot be used accurately when the perturbative in-in formalism fails to apply, highlighting the relevance of developing fully non-perturbative approaches to the problem.
	
	\begin{bottompar}
		\noindent\xdash[15em]\\
		\small{
			guillermo.ballesteros@uam.es, 
			j.gambin@csic.es, 
thomas.konstandin@desy.de,
alejandro.perezrodriguez@uam.es,\\
mathias.pierre@desy.de,
julian.rey@desy.de}
	\end{bottompar}
\end{titlepage}

\setcounter{page}{2}
\newpage

\tableofcontents

\section{Introduction}

In recent years, considerable effort has been devoted to understanding inflation beyond the slow-roll regime. Particular attention has been paid to models featuring a phase of ultra slow-roll (USR) \cite{Kinney:2005vj}, which is realized in single-field inflation models whose potential has a non-stationary inflection point, which makes the inflaton slow down. In this scenario, the Fourier modes of scalar perturbations exiting the horizon around the USR phase are enhanced, leading to a peak in the primordial power spectrum at those scales, which can eventually trigger the formation of a significant amount of primordial black holes (PBH) (see e.g.\ \cite{Garcia-Bellido:2017mdw, Ballesteros:2017fsr, Ballesteros:2020qam} for concrete examples of models). These PBH may account for the dark matter (see \cite{Green:2020jor} for a recent review on PBH bounds). Concurrently, the breakdown of slow-roll around the inflection point leads to non-Gaussianities that are also enhanced around the same scales. Since the PBH abundance is highly sensitive to the details of the probability distribution function (PDF) of curvature fluctuations at large values, it is important to determine it as accurately as possible. Indeed, there is growing interest on the characterization of the tail of said PDF and, more in general, on studying inflationary perturbations in the non-perturbative regime (see e.g. \cite{Ezquiaga:2019ftu,Figueroa:2020jkf,Celoria:2021vjw,Caravano:2024moy} for results from various perspectives).

In this paper, we focus on the non-Gaussianity arising in single-field inflation models with a phase of USR and consider its manifestation on the tail of the PDF of curvature fluctuations. Much of the literature on the topic,  specifically in the context of PBH formation, has followed one of two approaches:
\begin{itemize}
\item The computation of high-order (three or larger) correlators of $\zeta$ --the curvature perturbation on slices of uniform density--, which encode information about its non-Gaussian statistics around  $\zeta =0$. These calculations are usually performed in the $\zeta$-gauge, in which the only scalar degree of freedom is placed in the spatial part of the metric, see in particular \cite{Saito:2008em, Taoso:2021uvl} for PBH formation from USR. The possibility of extracting information about the tail of the PDF for large $\zeta$ from these correlators is hampered by their own definition. See \cite{Celoria:2021vjw} for a clear introductory discussion illustrating this difficulty. However, there are several works which have studied the effects of perturbative non-Gaussianities on the PBH abundance \cite{Franciolini:2018vbk,Atal:2018neu,DeLuca:2019qsy,Kalaja:2019uju}.
\item The computation of the non-Gaussian PDF of $\zeta$ by performing a non-linear transformation of the PDF of $\delta\phi$ (perturbation of the inflaton field), which is  \emph{assumed} to be Gaussian. The non-linear relation between the two variables can be obtained from the so-called $\delta N$ formalism \cite{Lyth:2004gb, Sugiyama:2012tj}, which provides the gauge transformation between $\zeta$ and $\delta\phi$ (in the $\delta\phi$-gauge, where the single scalar degree of freedom is placed in the matter field) to all orders in perturbations and second order in gradients, see also \cite{Ballesteros:2024zdp}.\footnote{This non-linear relation from the $\delta N$ formalism can also be implemented in the context of stochastic inflation \cite{Ezquiaga:2019ftu, Figueroa:2020jkf, Tomberg:2023kli}, rendering an equivalent result for the PDF of $\zeta$ from an USR phase \cite{Ballesteros:2024pwn} (due to the presence of an attractor and the fact that the power spectrum is peaked). See also \cite{Cruces:2024pni, Jackson:2024aoo}.} Examples of this approach are \cite{Atal:2019cdz, Atal:2019erb, Biagetti:2021eep, Hooshangi:2021ubn, Ferrante:2022mui, Pi:2022ysn, Hooshangi:2023kss}.
\end{itemize}

In this paper, we asses the validity of assuming a Gaussian $\delta\phi$ in the latter approach. This can be done by computing the bispectrum of $\delta\phi$ and higher-order correlators of this variable and using them to construct a non-Gaussian input for the non-linear relation between $\zeta$ and $\delta \phi$ in the $\delta N$ formalism.\footnote{This is different from what was done in \cite{Creminelli:2003iq}, where the bispectrum of $\delta\phi$ was used as input for a \emph{linear} relation with $\zeta$ to compute the bispectrum of the latter.} For simplicity, we restrict our analysis to including the effect of the bispectrum of $\delta\phi$, but our analysis can be generalized to include higher-order correlators as well. Our approach can be decomposed in three steps:	
\begin{enumerate}
\item We compute the tree-level power spectrum and bispectrum for $\delta \phi$ in the $\delta\phi$-gauge using the in-in formalism. This only requires knowing the Lagrangian for fluctuations of $\delta\phi$ at cubic order. As we mentioned above, our procedure can be generalized to include higher-order correlators. It can also accommodate loop corrections.
\item On a spatial lattice, we construct a random variable endowed with the spectrum and bispectrum computed for $\delta\phi$ in step 1. We sample the distribution of this random field (at the end of inflation) to get a proxy for the non-Gaussian PDF of $\delta\phi$.
\item We use the non-linear relation between $\delta\phi$ and $\zeta$ (encoded in the gauge transformation provided by the $\delta N$ formalism) to compute the non-Gaussian PDF of $\zeta$ from the one of $\delta\phi$ (obtained in the previous step). 
\end{enumerate}
We can distinguish two formally different contributions to the non-Gaussianity of $\zeta$. The first is the non-Gaussianity due to the {\it non-linear} relation between $\delta\phi$ and $\zeta$ (assuming Gaussian $\delta\phi$). The second is the correction over the former due to the non-Gaussianity of $\delta\phi$, coming from its self interactions, {which we account for perturbatively}. This correction will henceforth be referred to as {\it intrinsic}.\footnote{In the literature on PBH formation, the word {\it intrinsic} has been used earlier to refer to the non-Gaussianity in density fluctuations arising from their non-linear relation to curvature fluctuations, see e.g.\ \cite{DeLuca:2019qsy,Kehagias:2019eil,Yoo:2019pma}. That is a different notion from the one we are interested in in this work.} Of course, this distinction is artificial; if we were to compute the PDF of $\zeta$ in the $\zeta$-gauge, we would simply use the self interactions of $\zeta$. However, the distinction between the two contributions in the $\delta\phi$-gauge is convenient for two reasons. First, we are indeed interested in checking the commonly used assumption of Gaussianity for $\delta\phi$ in the derivation of the PDF of $\zeta$. Second, using the $\delta\phi$-gauge the calculations are simpler (especially for higher-order correlators and at higher loop order, see \cite{Ballesteros:2024zdp}). 

We consider an USR phase preceded by a standard slow-roll (SR) phase, followed by a constant-roll (CR) phase, characterized by a constant second slow roll parameter $\eta=\eta_\text{CR}< 0$ (which ensures that $\epsilon$ grows and inflation eventually ends). We parametrize the slow-roll parameters in such a way that the duration and smoothness of the USR phase can be controlled. 
We find that the bispectrum of $\delta\phi$ can introduce sizable corrections to the PDF of $\zeta$, depending on the duration of the phase and smoothness of the transitions in and out of it. However, in such a regime the approach may not allow to derive accurate predictions for the tail of the PDF. This is somewhat analogous to what happens with the one-loop corrections to the scalar power spectrum, which are also controlled by the cubic term in the Lagrangian for $\delta\phi$ \cite{Ballesteros:2024zdp}.

The paper is structured as follows. In Section \ref{sec:action}, we review the in-in formalism, discuss the cubic interactions of $\delta\phi$ in the $\delta\phi$-gauge and the gauge transformation to the $\zeta$-gauge. In Section \ref{sec:correlators}, we compute the bispectrum and use Maldacena's consistency relation as a check that we have included all relevant interactions. In Section \ref{sec:lattice}, we implement a numerical procedure to generate a random field in a lattice with the correct power spectrum and bispectrum, and determine the resulting PDF of $\zeta$. The conservation of the bispectrum of $\zeta$ on superhorizon scales is discussed in Appendix \ref{app:bispectrum}. In Appendix \ref{app:gram-charlier} we discuss how the Gram-Charlier series, which has sometimes been used to include perturbative non-Gaussian effects on the PDF of $\zeta$, fails in the present case.

\section{Fluctuations in ultra slow-roll inflation}
\label{sec:action}

\subsection{The inflationary setup}

Throughout the paper, we set $\hbar=c=1$, and we denote as $M_P = (8\pi\, G)^{-1/2}$ the reduced Planck mass. The starting point for our analysis is the canonical action for a minimally coupled scalar (inflaton) field $\Phi(t,\boldsymbol{x})$ with potential $V$,
\begin{equation}
S=\int \diff^4x\sqrt{-g}\bigg[\frac{M_P^2}{2}R-\frac{1}{2}g^{\mu\nu}\partial_\mu\Phi \partial_\nu\Phi-V(\Phi)\bigg]\,,
\end{equation}
where $R$ denotes the Ricci scalar. For the background (homogeneous) inflaton field $\phi(t)$, the metric reduces to FLRW, with scale factor $a(t)$. The (homogeneous) background equations of motion for the field and metric are
\begin{equation}
\label{eq:EoMphi}
\phi'' = -(3-\epsilon)(\phi'+M_P^2\,\partial_\phi V/V) , \qquad \, M_P^2\,H^2(3-\epsilon)=V\,,
\end{equation}
where we use the amount of expansion measured in e-folds, $N$ (defined  via   $\diff N = H \diff t = \dot a/a \diff t$) as time variable. Primes denote the corresponding derivatives, $^{\prime} \equiv \diff /\diff N$, and dots indicate standard (cosmic) time derivatives, $\dot{} \equiv \diff /\diff t$. The quantity $\epsilon$ appearing in Eq.\ \eqref{eq:EoMphi} is the first slow-roll parameter, whose definition is	
\begin{equation}
\epsilon\equiv -\frac{\dot{H}}{H^2}=\frac{1}{2M_P^2}\frac{\dot{\phi}^2}{H^2}\,.
\end{equation}
For convenience, we also introduce the second slow-roll parameter
\begin{align}
\eta\equiv -\frac{1}{2}\frac{\dot{\epsilon}}{H\epsilon}\,,
\end{align}
which measures the acceleration of the background inflaton field. 	

We are interested in inflationary models with a potential featuring a non-stationary inflection point. In these models, the inflaton goes first through a SR phase (with $\epsilon, |\eta| \ll 1$).\footnote{The large-scale fluctuations observed imprinted in the CMB correspond to this phase in these models.}  This stage ends as the inflaton approaches the local minimum, and is followed by an USR phase (with $\epsilon \ll 1$ and $\eta\geq 3/2$) in which the inflaton slows down as it climbs up the local maximum of the potential. Finally, the dynamics undergo a CR phase, characterized by a constant $\eta$, as the inflaton rolls down the local maximum towards the absolute minimum of the potential. During this last phase, the first slow-roll parameter $\epsilon$ grows until inflation ends. The slow-roll parameter $\epsilon$ remains small throughout the entire evolution and becomes of $\mathcal{O}(1)$ only at the very end of inflation.

We study perturbations to the background solution using the ADM formalism \cite{Arnowitt:1962hi}. The metric is parametrized as
\begin{equation}
\diff s^2=-\mathscr{N}^2 \diff t^2+h_{ij}\big(\diff x^i+\mathscr{N}^i \diff t\big) \big(\diff x^j+\mathscr{N}^j \diff t \big),
\end{equation}
where the lapse $\mathscr{N}$ and shift $\mathscr{N}_i$ are Lagrange multipliers that must be determined via their (non-dynamical) equations of motion and inserted back into the action. To perform the calculation, it is convenient to fix a gauge. Let us split the field into a homogeneous piece and a perturbation
\begin{equation}
\Phi(t,{\bm x})=\phi(t)+\delta\phi(t,{\bm x}).
\end{equation}
We can either work in the $\zeta$-gauge, defined by the conditions
\begin{equation} \label{eq:zetagauge}
\delta\phi(t,{\bm x})=0,\qquad h_{ij}=a^2e^{2\zeta}\delta_{ij},
\end{equation}
or in the $\delta\phi$-gauge, defined by
\begin{equation}
\zeta(t,{\bm x})=0,\qquad h_{ij}=a^2\delta_{ij}.
\end{equation}
In this way, by using a coordinate transformation, we place the only scalar degree of freedom in the theory either in the metric or in the matter sector, respectively. Throughout the rest of this work, we neglect the effect of vector and tensor perturbations. The full action for the perturbations in either gauge contains, prior to any approximation, multiple interaction terms at any order, due to the non-linearities introduced by the lapse and shift. However, since the first slow-roll parameter $\epsilon$ is small during ultra slow-roll --in fact, it remains small until inflation approaches its end-- and the second parameter $\eta$ is of $\mathcal{O}(1)$ at most and many interaction terms are proportional to powers of $\epsilon$, the action simplifies considerably for the kind of models we are interested in. This is more evident in the $\delta\phi$-gauge, which is also convenient, since our goal is to explore the relevance of the non-Gaussianity of $\delta\phi$.

\subsection{The in-in formalism} \label{sec:In-in}

We will use the in-in formalism \cite{Weinberg:2005vy} --see also  \cite{Chen:2010xka, Wang:2013zva} and Appendix A of  \cite{Ballesteros:2024zdp} for reviews of various lengths-- to compute the bispectrum of $\delta\phi$. Here we summarize a few basic equations of the formalism that will be useful later on.

Given a general Hermitian operator\footnote{{We will denote quantum operators and quantized fields with ``hats'', $\,\hat{\,}$.}} $\mathcal{\hat O}(t)$ in the Heisenberg picture, its vacuum expectation value is:
\begin{equation}\label{ininbasic}
	\expval{\mathcal{\hat O}(t)} = \bra{0} \hat F^{-1}(t,-\infty_+) \mathcal{\hat O}_I(t)\hat F(t,-\infty_-) \ket{0} {\bigg |}_\textrm{no bubbles}\,,
\end{equation}
where the subscript $_I$ carried by the operator $\mathcal{\hat O}_I(t)$ denotes the interaction picture and $\ket{0}$ represents the interaction vacuum.\footnote{In the Heisenberg picture, the vacuum state does not evolve in time, but the operators, such as $\mathcal{\hat O}(t)$ do evolve. In the interaction picture the fields evolve in time according to the free Hamiltonian (without interactions) and satisfy standard commutation relations; and the corresponding vacuum is the one of the free Hamiltonian. In the derivation of the in-in formula \eq{ininbasic} one starts with the Heisenberg picture and then expresses fields and vacuum in that picture in terms of the ones in the interaction picture, see Appendix A of  \cite{Ballesteros:2024zdp} for details.} The time evolution operator in the interaction picture is:
\begin{equation}
	\hat{F}(t,t_0) = T \exp \left( -i \int_{t_0}^t \dd t' \hat{H}_I(t') \right) \quad {\rm and} \quad \hat{F}^{-1}(t,t_0) = \overline{T} \exp \left(i \int_{t_0}^t \dd t' \hat{H}_I(t') \right)\,,
\end{equation}
where $T$ and $\overline{T}$ denote time-ordering and anti time-ordering, respectively. We stress that $\expval{\hat{\mathcal{O}}(t)}$ does not receive bubble contributions \cite{Ballesteros:2024qqx}, so all insertions of $\hat{H}_I$ must be connected to the operator $\hat{\mathcal{O}}_I$.
In practice, we will only need to calculate expectation values with a single Hamiltonian insertion (since we are interested in the bispectrum of $\delta\phi$), so that
\begin{equation}
	\expval{\hat{\mathcal{O}}(t)} = \bra{0} \hat{\mathcal{O}}_I(t) \ket{0} +2\, {\rm Im} \left\lbrace  \int_{-\infty_-}^t \dd t'\, \bra{0} \hat{\mathcal{O}}_I(t) \hat{H}_I(t') \ket{0}\right\rbrace + \mathcal{O}\left( \hat{H}_I ^2\right) \,.
\end{equation}
The prescription $i\, \omega$ (with an infinitesimal $\omega>0$ that eventually will be sent to $0$) in the lower limit of time integrals, $-\infty_{\pm} \equiv -\infty(1\pm i\omega)$, ensures that the system projects onto the interaction picture vacuum $\ket{0}$ at times $t \to -\infty$. The operators $\hat{\mathcal{O}}_I$ and $\hat{H}_I$, in the interaction picture, are composed of fields that evolve following the dynamics governed by the free Hamiltonian.
Therefore, these fields can be decomposed into creation and annihilation operators as usual. 
For the fluctuations of the inflaton in this picture, we have
\begin{equation}\label{ec:quant_free}
	\hat{\delta\phi} (t,\boldsymbol{x}) =  \int \dfrac{\dd^3 k}{(2\pi)^{3}} e^{i \bm{k} \bm{x}} \hat{\delta\phi}_{\bm{k}} (t)\,, \quad {\rm where} \quad \hat{\delta\phi}_{\bm{k}} (t) = \delta \phi_k(t) \hat a_{\bm{k}} 
	+ \delta \phi^* _k(t) \hat a^\dagger_{-\bm{k}}  \,.
\end{equation}
The creation and annihilation operators, $\hat a_{\bm{k}}^\dagger$ and $\hat a_{\bm{k}}$, satisfy $\left[\hat a_{\bm{k}},\, \hat a^\dagger_{\bm{p}} \right] = {(2\pi)^3} \delta \left(\bm{k} - \bm{p} \right) $.
The Fourier modes $\delta\phi_k(t)$ satisfy the free equation of motion which, in this case, at first order in the SR parameter $\epsilon$, is
\begin{equation}
	\ddot{\delta \phi}_{k} + 3 H \dot{\delta\phi}_{k} + \left( \dfrac{k^2}{a^2}+ \partial_\phi^2V\right) \delta \phi_{k} = 0 \,,
	\label{eq:deltaphi_2}
\end{equation}
	where, neglecting terms suppressed by $\epsilon$, the second derivative of the potential can be written as
	\begin{equation}
	\partial_\phi^2V\simeq -\frac{H^2}{2}\left(3\eta+\frac{\eta^2}{2}+\frac{\dot\eta}{H}\right)\,.
	\end{equation}
Besides, $\delta\phi_k(t)$ satisfies Bunch-Davies initial conditions (which ensure canonical commutation relations):
\begin{equation}
	 \lim_{\tau \to -\infty} \delta\phi_k(\tau) = \frac{1}{a(\tau)} \frac{e^{-i k \tau}}{\sqrt{2 k}}\,,
\end{equation}
where the factor $1/a$ comes from the canonical normalization of $\delta\phi$, and $\tau$ is the conformal time defined according to $\dd t = a(\tau) \dd \tau$.

The $n$-point correlation functions in Fourier space are defined as
\begin{equation}
	\expval{\hat{\delta\phi}(t,\bm{x}_1)\cdots \hat{\delta\phi}(t,\bm{x}_n)} \equiv \int \prod_{j = 1}^n \left( \dfrac{\dd^3 k_j}{(2\pi)^{3}} e^{i \bm{k}_j \bm{x}_j} \right) \expval{\hat{\delta\phi}_{\bm{k}_1}(t)\cdots \hat{\delta\phi}_{\bm{k}_n}(t)}\,.
\end{equation}
The connected two-point correlation is related to the dimensionless power spectrum as follows:
\begin{equation}\label{ec:power_spectrum_1}
\mathcal{P}_{\delta\phi}(t,\bm{k})\,(2\pi)^3\,\delta(\bm{k}+\bm{k}') \equiv \frac{k^3}{2\pi^2}\langle\delta\hat{\phi}_{\bm{k}}(t)\,\delta\hat{\phi}_{\bm{k}'}(t)\rangle_c\,.
\end{equation}
The subscript $_c$ refers to the connected part of the correlation, defined as the contribution having a single Dirac delta that guarantees the momentum conservation.

\subsection{The $\delta N$ formula} \label{sec:delta N formula}

We now discuss the relation between the scalar degrees of freedom in the $\delta\phi$-gauge and the $\zeta$-gauge. For this purpose, it is necessary to consider a transformation connecting both gauges.
Neglecting gradients and at late times,\footnote{Neglecting gradients implies that the time $t$ must be chosen once the USR phase is over and the Fourier modes of interest for the problem we are studying (which defines a certain coarse-graining scale) are super-Hubble.} the perturbation $\zeta$ (defined in the $\zeta$-gauge, see Eq.\ \eqref{eq:zetagauge}) can be expressed as a perturbation in the local expansion rate \cite{Ballesteros:2024zdp}
\begin{equation}
\zeta=\log\left[\frac{a(t+\delta t)}{a(t)}\right],
\label{eq:zeta_log}
\end{equation}
where the infinitesimal time transformation $\delta t$ is the one connecting the $\delta\phi$-gauge (left-hand side) and the $\zeta$-gauge (right-hand side):
\begin{equation}
\phi(t+\delta t)+\delta\phi(t+\delta t)=\phi(t)\,.
\label{eq:field_trans}
\end{equation}
Using Eq.\ \eqref{eq:field_trans}, we can find $\delta t$ as a function of $\delta\phi$, and then use \eqref{eq:zeta_log} to determine $\zeta$ as a function of $\delta\phi$ up to gradients and order by order in perturbations
\begin{equation}
\zeta
=
-\frac{H}{\dot{\phi}}\delta\phi
-\frac{\eta}{2}\bigg(-\frac{H}{\dot{\phi}}\delta\phi\bigg)^2
+\frac{1}{3}\bigg(\eta^2+\frac{\dot{\eta}}{2H}\bigg)\bigg(-\frac{H}{\dot{\phi}}\delta\phi\bigg)^3+\mathcal{O}(\delta\phi^4).
\label{eq:r_nonlin}
\end{equation}
To get to this expression, we have used the fact that, neglecting gradients and up to second order in perturbations, the equation of motion for $\delta\phi$ reads \cite{Ballesteros:2024zdp}
\begin{equation}
\frac{\diff }{\diff t}\bigg[
\bigg(-\frac{H}{\dot{\phi}}\delta\phi\bigg)
-\frac{\eta}{2}\bigg(-\frac{H}{\dot{\phi}}\delta\phi\bigg)^2\bigg]=\mathcal{O}(\delta\phi^3).
\end{equation}
Incidentally, this equation makes manifest the conservation of $\zeta$ on super-Hubble scales ($k\ll aH$) up to second order in perturbations (see \cite{Weinberg:2003sw} for a proof to first order, and \cite{Maldacena:2002vr, Lyth:2004gb} for a proof up to all orders).

It should be noted that \eqref{eq:zeta_log} and \eqref{eq:field_trans} are nothing but the $\delta N$ formula, see \cite{Lyth:2004gb}, which non-linearly relates (on super-Hubble scales) the curvature perturbation on uniform density slices with the difference in local expansion due to a field perturbation on spatially-flat slices. This \emph{$\delta N$ formalism} allows to compute \eqref{eq:r_nonlin} to all orders in $\delta\phi$ by just solving the background equations of motion for $\phi$ and $H$ with shifted initial conditions.

For the particular case in which $\eta$ is approximately constant (as it is the case in the CR phase after USR, when the Fourier modes in the peak of the curvature power spectrum freeze out), the background equations of motion can be solved analytically, and the $\delta N$ formula has a closed form \cite{Atal:2019cdz, Atal:2019erb, Tomberg:2023kli, Ballesteros:2024pwn}  
\begin{equation}
\eta_{\rm CR}\zeta=\log\bigg(1-\eta_{\rm CR}H\frac{\delta\phi}{\dot{\phi}}\bigg)\,,
\label{eq:deltan_cr}
\end{equation}
where $\eta_{\rm CR}$ is the constant value of $\eta$ along the CR attractor. As a consistency check, notice that expanding \eqref{eq:deltan_cr} as a Taylor series around $\delta\phi=0$, one recovers \eqref{eq:r_nonlin} for constant $\eta$.

If one \emph{assumes} $\delta\phi$ to be Gaussian, the PDF of $\zeta$ obtained from \eqref{eq:deltan_cr} has the usual exponential tail reported for USR models (see \cite{Ballesteros:2024pwn} for a recent discussion),
\begin{equation}
P(\zeta)=\frac{1}{\sigma\sqrt{2\pi}}\exp
\bigg[
-\frac{1}{2\sigma^2\eta_{\rm CR}^2}
\Big(1-e^{\eta_{\rm CR}\zeta}\Big)^2
+\eta_{\rm CR}\zeta
\bigg],
\label{eq:pdf_r}
\end{equation}
where
\begin{equation}\label{ec:variance}
\sigma^2=\frac{H^2}{\dot{\phi}^2}\int\mathcal{P}_{\delta\phi}(k)\diff \log k
\end{equation}
is the variance of $\zeta$ in linear perturbation theory. In the integrand, $\mathcal{P} _{\delta\phi}$ is the dimensionless power spectrum of $\delta\phi$ defined in \eqref{ec:power_spectrum_1} which, at tree level, is
\begin{equation}
	\mathcal{P}_{\delta\phi}({k}) = \frac{k^3}{2\pi^2}|\delta\phi_k|^2\,.
\end{equation}
To complete the description of the PDF of $\zeta$, it is necessary to analyze which are the integration limits that appear in Eq.\ \eqref{ec:variance}. These limits, or cutoffs, are dependent on the problem under consideration and are closely related to the observable we are interested in. In our case, as we have been discussing, we are interested in PBH formation in post-inflationary epochs; a process that occurs when a $\zeta$-mode enters the horizon with sufficient amplitude to generate a matter fluctuation that eventually collapses forming a PBH.
To characterize the PDF of $\zeta$ in such a way that it captures the information associated with the PBH formation, we must choose the cutoffs in a way that resolves the physics related to the USR phase during inflation, i.e.\ we must capture the peak of the power spectrum in the variance, Eq.\ \eqref{ec:variance}. 
This is because at these scales the probability of generating a fluctuation of $\zeta$ large enough to end up forming a PBH is maximal, since it is precisely on these scales that the power spectrum reaches its highest value.\footnote{There are different prescriptions to compute the PBH abundance --see e.g.\ \cite{Young:2024jsu} for a recent discussion-- but all involve setting cutoffs via window functions. The specific prescription that is used (e.g.\ Press-Schechter vs peak theory) leads to quantitative differences in the PBH abundance, but it is not relevant for our purpose, i.e.\ illustrating the relevance of non-Gaussianities in the PDF of $\zeta$. These different approaches will not give qualitatively significant differences as long as the peak of the power spectrum lies inside the chosen cutoffs.}

We are interested in trying to determine the deviation from \eqref{eq:pdf_r} induced by the fact that $\delta\phi$ is not exactly a free field. {In the following sections,} we will study this issue by considering only the three-point function of $\delta\phi$.

\subsection{Relevant interactions}

We now determine the interaction terms relevant for the calculation of the bispectrum of $\delta\phi$ in the eponymous gauge.
It is convenient to define the rescaled inflaton perturbation\footnote{This is a definition, there is no truncation in powers of $\delta\phi$. This variable coincides with the linear term (in powers of $\delta\phi$) of the comoving curvature perturbation in the flat gauge. It also coincides with the redefined field $\zeta_n$ introduced in \cite{Maldacena:2002vr} and used in \cite{Taoso:2021uvl}. We emphasize that \eqref{ec:varphi} implies no change of gauge: $\varphi$ is the single scalar degree of freedom in the $\delta\phi$-gauge.\label{ft:equivalence_variables}} 
\begin{equation}\label{ec:varphi}
\varphi\equiv-\frac{H}{\dot{\phi}}\delta\phi\,.
\end{equation}
The interaction Lagrangian at cubic order in the $\delta\phi$-gauge is \cite{Maldacena:2002vr} \nopagebreak
\begin{align} \nonumber
\frac{\mathcal{L}_I^{(3)}}{M_P^2}
=&
-c_0\, a^3H^2\varphi^3
-c_1\, a^3\varphi\,\dot{\varphi}^2
-c_2\, a^3\dot{\varphi}\,\partial_i\varphi\,\partial^i\left(\partial^{-2}\dot{\varphi}\right)
-c_3\, a\,\varphi\,\left(\partial_i\varphi\right)^2
\\& 
-c_4\, a^3
\varphi\left(\partial_i\partial_j\left(\partial^{-2}\dot{\varphi}\right)\right)^2
+\frac{\diff}{\diff t}\left(c_5\, a^3H\varphi^3\right),
\label{eq:fullL}
\end{align}
where the coefficients are given by 
\begin{equation} \label{c0}
c_0= \frac{1}{3}\epsilon\,\eta\,\epsilon_3(\epsilon+2\eta-\epsilon_3-\epsilon_4-3)
\end{equation}
and
\begin{equation}
c_1= \frac{\epsilon^2}{2}(\epsilon-2)\,,\quad
c_2= 2\epsilon^2\,,\quad
c_3 = -\epsilon^2\,,\quad
c_4= -\frac{\epsilon^3}{2}\,,\quad
c_5= -\frac{\epsilon^2}{3}(\epsilon+3\eta).
\label{eq:smallc}
\end{equation}
We define recursively $\epsilon_n = \dot{\epsilon}_{n-1}/(H\epsilon_{n-1})$, with $\epsilon_1=\epsilon$ and $\epsilon_2=-2\eta$. One can see that all the $c_n$ coefficients with $n\geq 1$ are suppressed by powers of $\epsilon \ll 1$ with respect to $c_0$. Similarly, during an USR phase, $\eta$ changes quickly and $\epsilon_3$ can grow significantly, so that the only relevant term in the presence of an USR phase is the inflaton self interaction $c_0\,\varphi^3$. This term contains all contributions to $\mathcal{L}_I^{(3)}$ that are proportional to $\varphi^3$, and also those proportional to $\varphi^2\dot{\varphi}$ after integrating by parts in time. Within these contributions, the main one is the one coming from the third order in the Taylor expansion of the potential, which is proportional to $\partial_\phi^3V$, and becomes large around the inflection point of the potential. From now on, we neglect the rest of the interactions. Therefore, we approximate the interaction Hamiltonian in the interaction picture, in the $\delta\phi$-gauge, as 
\begin{equation}
\hat H_I(t)
=
M_P^2\int \diff^3{\bm x}\,
c_0(t) a(t)^3H(t)^2
\hat \varphi(t,{\bm x})^3\,,
\label{eq:cubic_hamiltonian}
\end{equation}
where the scalar fluctuation $\hat{\varphi}$ in the interaction picture is:
\begin{equation}\label{eq:varphi int pic}
	\hat{\varphi}(t,\bm{x}) = \int \dfrac{\dd^3 k}{(2\pi)^3} e^{i\bm{k} \bm{x}} \hat{\varphi}_{\bm{k}}(t)\,, \quad {\rm where} \quad \hat{\varphi}_{\bm{k}}(t) =  \varphi_k(t) \hat{a}_{\bm{k}} + \varphi^*_k (t) \hat{a}_{-\bm{k}}^\dagger \,.
\end{equation}
The free modes satisfy the equation of motion
\begin{equation}\label{ec:MS}
	\varphi''_k+(3+\epsilon-2\eta)\varphi_k'+\left(\frac{k}{aH}\right)^2\varphi_k=0\,,
\end{equation}
with the initial conditions determined by Bunch-Davies, after canonical normalization.

\section{Bispectrum of $\zeta$}

\label{sec:correlators}

{In this section, we compute the lowest order contribution to the bispectrum of $\zeta$ in an expansion on $H/M_P$ in order to obtain some insight into whether the (perturbative) non-Gaussianity of $\delta\phi$ can yield an important contribution to that of $\zeta$ at values of $\zeta$ relevant for PBH formation. The intuition developed with this calculation will help us to interpret the results in Sec.~\ref{sec:lattice} for the non-Gaussian PDF of $\zeta$ including corrections arising from the non-Gaussianity of $\delta\phi$}. 

\subsection{Contributions to the bispectrum of $\zeta$}

During the CR stage that follows the USR phase, the curvature perturbation is related to $\varphi$ as
\begin{equation}
 \zeta  \, = \, \varphi - \dfrac{\eta_\text{CR}}{2} \varphi^2 + \dfrac{\eta_\text{CR}^2}{3} \varphi^3 + \cdots.
    \label{eq:expansionR}
\end{equation}
This can be obtained by evaluating \eqref{eq:r_nonlin} for constant $\eta=\eta_{\rm CR}$, or, equivalently, expanding \eqref{eq:deltan_cr} around $\delta\phi=0$. We now consider the Fourier modes $\hat{\zeta}_{\bm k}$ of the quantum operator $\hat{\zeta}$, related to $\hat{\varphi}$ by \eqref{eq:expansionR}. Expanding the three-point correlation function of $\hat{\zeta}$ in powers of $\hat{\varphi}$, the two lowest order terms in $H/M_P$ are 
\begin{equation}
\langle\hat{\zeta}_{\bm p}\hat{\zeta}_{\bm q}\hat{\zeta}_{\bm k}\rangle
= \langle\hat{\varphi}_{\bm p}\hat{\varphi}_{\bm q}\hat{\varphi}_{\bm k}\rangle - \dfrac{\eta_\text{CR}}{2}  \bigg( \int \frac{\diff^{3} \ell}{(2 \pi)^{3}}\langle \hat \varphi_{\bm k-\bm \ell} \hat \varphi_{\bm \ell} \hat \varphi_{\bm q} \hat \varphi_{\bm p}\rangle + \text{perm.} \bigg)\,.
\label{eq:zetaexpansion}
\end{equation}
Introducing the notation
\begin{align}
\mathcal{I}({\bm p},{\bm q},{\bm k})(2\pi)^3\delta^{(3)}({\bm p}+{\bm q}+{\bm k}) &= \langle\hat{\varphi}_{\bm p}\hat{\varphi}_{\bm q}\hat{\varphi}_{\bm k}\rangle_c\,, \label{ec:I}\\
\mathcal{N}({\bm p},{\bm q},{\bm k})(2\pi)^3\delta^{(3)}({\bm p}+{\bm q}+{\bm k}) &= - \dfrac{\eta_\text{CR}}{2}  \left( \int \frac{\diff^{3} \ell}{(2 \pi)^{3}}\langle \hat \varphi_{\bm k-\bm \ell} \hat \varphi_{\bm \ell} \hat \varphi_{\bm q} \hat \varphi_{\bm p}\rangle_c + \text{perm.}\right)\,,\label{ec:N}
\end{align}
we obtain that the bispectrum of $\zeta$ is
\begin{equation}
\langle\hat{\zeta}_{\bm p}\hat{\zeta}_{\bm q}\hat{\zeta}_{\bm k}\rangle_c	 =B({\bm p},{\bm q},{\bm k}) (2\pi)^3\delta^{(3)}({\bm p}+{\bm q}+{\bm k})
= 
\Big[
\mathcal{I}({\bm p},{\bm q},{\bm k})
+\mathcal{N}({\bm p},{\bm q},{\bm k})
\Big](2\pi)^3\delta^{(3)}({\bm p}+{\bm q}+{\bm k})\,.
\end{equation}
If $\delta\phi$ had no interactions, $\mathcal{I}$ would be zero (since it corresponds the correlator of an odd number of fields). Its leading contribution arises from the tree-level diagram featuring a single insertion of the cubic interaction Hamiltonian:
\begin{equation}
\langle\hat{\varphi}_{\bm p}(t)\hat{\varphi}_{\bm q}(t)\hat{\varphi}_{\bm k}(t)\rangle_c
=
2\,{\rm Im}
\bigg[
\int_{-\infty_-}^t \diff t'
\langle
\hat{\varphi}_{\bm p}(t)\hat{\varphi}_{\bm q}(t)\hat{\varphi}_{\bm k}(t)
\hat{H}_I(t')
\rangle_c
\bigg].
\label{eq:threepointdeltaphi}
\end{equation}
Plugging in $\hat{H}_I$ from \eqref{eq:cubic_hamiltonian} and the field $\hat{\varphi}$ in the interaction picture described in Eq.\ \eqref{eq:varphi int pic}, we have
\begin{equation}
\mathcal{I}({\bm p},{\bm q},{\bm k})
=
3!M_P^2
\int_{-\infty_-}^t \diff t' a(t')^3c_0(t') H(t^\prime)^2
\;2\,{\rm Im}\Big[\varphi_p(t)\varphi_p^*(t')
\varphi_q(t)\varphi_q^*(t')
\varphi_k(t)\varphi_k^*(t')\Big]\,,
\label{eq:three_point}
\end{equation}
where we stress that $\varphi_k$ are the free modes satisfying \eqref{ec:MS}.
The contribution $\mathcal{N}$ in \eqref{ec:N} is non-zero even if $\delta\phi$ is non-interacting, since it is a correlator of an even number of fields. Therefore, its leading contribution is obtained by directly applying Wick's theorem, and corresponds to the {\it non-linear} contribution to the bispectrum. The result reduces to permutations of expectation values in the free vacuum of pairs of free fields \cite{Taoso:2021uvl},
\begin{equation}\label{eq:calN}
\mathcal{N}({\bm p},{\bm q},{\bm k}) = -\eta_{\rm CR}
\Big(
|\varphi_q|^2|\varphi_k|^2
+|\varphi_p|^2|\varphi_k|^2
+|\varphi_q|^2|\varphi_p|^2
\Big).
\end{equation}

To recap, we have obtained simple expressions for the lowest order contributions, $\mathcal{I}$ and $\mathcal{N}$, to the bispectrum of ${\zeta}$. On the one hand, $\mathcal{I}$ is given by the tree-level bispectrum of $\varphi$, and it accounts for the cubic self interactions of $\varphi$. In particular, it is zero for a free (i.e.\ Gaussian) $\varphi$; in other words, it gives an \emph{intrinsic} non-Gaussian contribution to the bispectrum of $\zeta$. On the other hand, $\mathcal{N}$ arises because of considering a {\it non-linear} (in this truncated calculation, quadratic) relation between $\delta\phi$ and $\zeta$. It is non-zero even if $\varphi$ is a free (i.e.\ Gaussian) field. At the lowest order in $H/M_P$, the split between intrinsic non-Gaussiniaty and that coming from non-linearities is manifest for the bispectrum. Indeed, the total bispectrum of $\zeta$ is the sum of \eqref{eq:three_point}, arising solely from $\delta\phi$ interactions, and \eqref{eq:calN}, due to non-linearity, which contains no information on $\delta\phi$ interactions. At higher orders in $H/M_P$, however, this separation is not so transparent. For example, if one computes loop corrections to \eqref{eq:calN}, one needs to include insertions of the cubic (or higher order) interaction Hamiltonian to compute the correlator in \eqref{ec:N}, which partly accounts for the intrinsic non-Gaussianity in the bispectrum of $\zeta$. 

A priori, it is not necessary to stop at the quadratic order in the non-linear relation between $\delta\phi$ and $\zeta$. The $\delta N$ formalism provides a non-linear relation between them, of which \eqref{eq:expansionR} is a truncated expansion (whose first non-linear term gave  us $\mathcal{N}$). It is however instructive to use the formulae computed in this section to gain some intuition on the relative contribution to the non-Gaussianity of $\zeta$ of the non-Gaussianity of $\delta\phi$ by simply comparing the size of $\mathcal{I}$ and $\mathcal{N}$. {One might naively think that the intrinsic contribution to the bispectrum of $\zeta$, i.e.\ $\mathcal{I}$, should always be smaller than the one coming from the change of variables, i.e\ $\mathcal{N}$, given that $\mathcal{I}$ is proportional to $c_0$ (which is suppressed by $\epsilon\ll 1$) and $\mathcal{N}$ comes with a factor $\eta_{\rm CR}$ (see Eq.\ \eqref{ec:N}), which can be $\mathcal{O}(1)$. However, this is actually not the case. A quick estimate of the relative importance between $\mathcal{I}$ and $\mathcal{N}$ can be obtained from a more careful analysis of the quotient of the slow-roll parameters involved in Eqs.\ \eqref{ec:I} and \eqref{ec:N}. Each $\varphi$ comes with a factor $\epsilon^{-1/2}$ from Eq.~\eqref{ec:N}. Besides, in Eq.~\eqref{ec:I} $c_0\sim \epsilon\,\eta\,\epsilon_3$ (from Eq.\ \eqref{c0}), while the prefactor in Eq.\ \eqref{ec:N} is directly $\eta_{\rm CR}$, as we already mentioned. Thus, we have $\mathcal{I}/\mathcal{N}\sim \epsilon_3 =(\log \epsilon_2)'$, which can be large depending on the sharpness of the transition. Moreover, the integration in time in Eq.\ \eqref{ec:I} (made explicit in Eq.\ \eqref{eq:three_point}) can alter the naive prediction on the hierarchy between $\mathcal{I}$ and $\mathcal{N}$ obtained by just counting slow-roll parameters. We will do a numerical comparison of the sizes of $\mathcal{I}$ and $\mathcal{N}$ in Section \ref{sec:num_model}, discussing the impact of the smoothness of the transitions between SR, USR and CR.}

\subsection{Consistency relation}

\begin{figure}
\centering
\includegraphics[width=1.0 \textwidth]{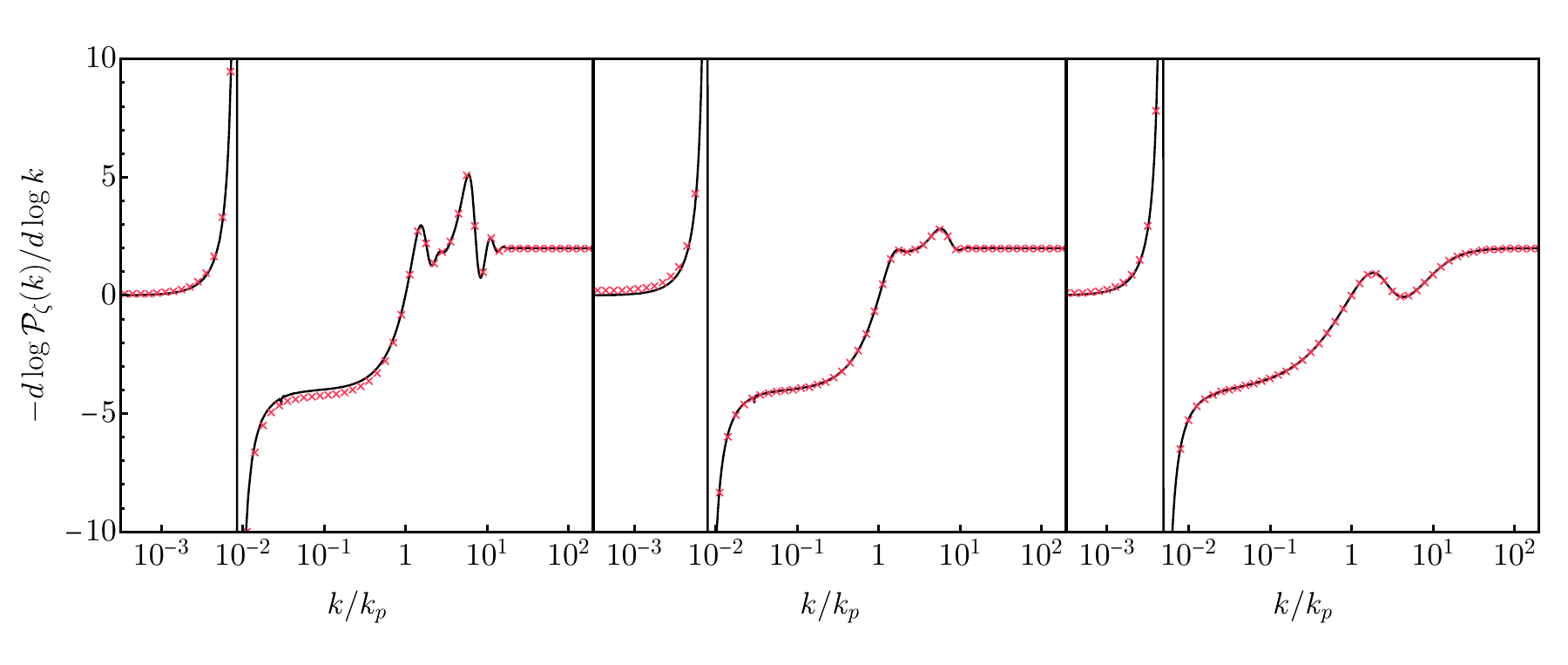}
\caption{
\it Consistency relation for the sharp (left), intermediate (center) and smooth (right) cases in Table \ref{table:models}, obtained by comparing the left-hand side of Eq.\,(\ref{eq:consistency}) (red crosses) and the right-hand side involving the derivative of the power spectrum (solid black line). The agreement implies that we have considered all interactions that are relevant (at least) in the squeezed limit.
}
\label{fig:consistency_I}
\end{figure}

Before moving on, we indicate a quick consistency check that one can make to detect whether any term neglected in \eqref{eq:cubic_hamiltonian} (and therefore in the calculation of $\mathcal{I}$) was actually non-negligible. We use the consistency relation for the bispectrum in the squeezed limit \cite{Maldacena:2002vr,Creminelli:2004yq}. If we take $k\ll p\simeq q$ in Eq.\,(\ref{eq:three_point}) and use the fact that $\varphi_k$ freezes in the super-Hubble limit, we find\footnote{One might worry that, since the time integral starts at $t'\to-\infty_-$, there is some integration range in which $\varphi_k(t')$ is not constant, no matter how early it exited the horizon. However, the fact that $\delta\phi$ self interactions are localized around the USR phase (in other words, that $c_0$ quickly decays away from the USR phase) make it safe to restrict the integration domain to a neighbourhood on said phase, which makes it a good approximation to take $\varphi_k(t')$ as constant as long as it exited the horizon sufficiently before USR. }
\begin{equation}
\frac{\mathcal{I}(k,p)}{|\varphi_k(t)|^2|\varphi_p(t)|^2}
=
12M_P^2\,{\rm Im}\bigg[\frac{\varphi_p(t)^2}{|\varphi_p(t)|^2}
\int_{-\infty_-}^t \diff t' a(t')^3c_0(t') H(t^\prime)^2
\varphi_p^*(t')^2
\bigg].
\end{equation}
Similarly,
\begin{equation}
\frac{\mathcal{N}(k,p)}{|\varphi_k(t)|^2|\varphi_p(t)|^2}
=
-\eta_{\rm CR}
\bigg(
2+\frac{k^3}{p^3}\frac{\mathcal{P}_\zeta(p)}{\mathcal{P}_\zeta(k)}
\bigg)
\simeq -2\eta_{\rm CR},
\end{equation}
where $\mathcal{P}_\zeta(k)= (k^3/2\pi^2)|\varphi_k|^2$ is the tree level power spectrum of $\zeta$. The following relation must be obeyed by virtue of the consistency relation:
\begin{equation}
\frac{\mathcal{I}(k,p)+\mathcal{N}(k,p)}{|\varphi_k(t_e)|^2|\varphi_p(t_e)|^2}
=
-\frac{\diff \log \mathcal{P}_\zeta(p)}{\diff \log p},
\label{eq:consistency}
\end{equation}
where $t_e$ denotes the end of inflation. If, while using our simplified $\hat{H}_I$ to compute $\mathcal{I}$, Eq.\ \eqref{eq:consistency} did not hold, it would mean that some non-negligible term from \eqref{eq:fullL} was not accounted for in \eqref{eq:cubic_hamiltonian}. We do this check in  Figure \ref{fig:consistency_I} for the model described in the next subsection. As shown in the figure, the equality of both sides of \eqref{eq:consistency} holds. Although this check is encouraging, we note that some interactions become negligible precisely in the squeezed limit (see Appendix \ref{app:bispectrum}) and this check is not sensitive to these contributions. See~\cite{Kristiano:2023scm} and~\cite{Motohashi:2023syh} for previous checks of the consistency relation in related settings. It is also worth recalling that earlier works pointed out that the consistency relation does not hold if the power spectrum and the bispectrum are evaluated at a time at which $\zeta$ has not frozen in the super-Hubble limit, see for instance \cite{Namjoo:2012aa,Martin:2012pe} for the specific case of USR inflation. This can be expected, given that the most straightforward derivation of the consistency relation relies explicitly on the conservation of $\zeta$, see \cite{Maldacena:2002vr}. A generalization of the consistency relation for the case in which the modes of $\zeta$ do not freeze in the super-Hubble limit was proposed in \cite{Bravo:2017wyw}. Later works explored the impact on the bispectrum (and on the consistency relation) of a transition from USR to SR, see in particular \cite{Cai:2018dkf, Passaglia:2018ixg} and also reference \cite{Taoso:2021uvl}, which we have already mentioned. As we have explained, we use the consistency relation as a tool aiming to test that we are including the necessary interactions in our computation of the bispectrum and therefore we check it at late times, when it has to apply in the type of model we consider.

\subsection{Numerical model}
\label{sec:num_model}

\begin{figure}
\centering
\includegraphics[width=0.49 \textwidth]{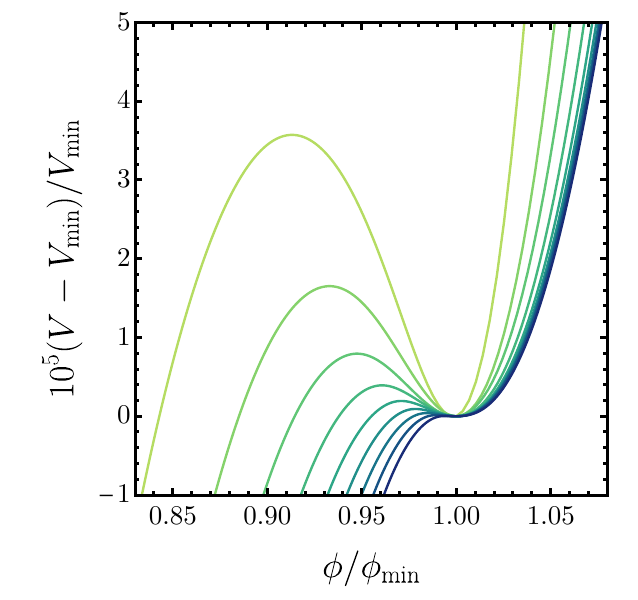} 
\includegraphics[width=0.49 \textwidth]{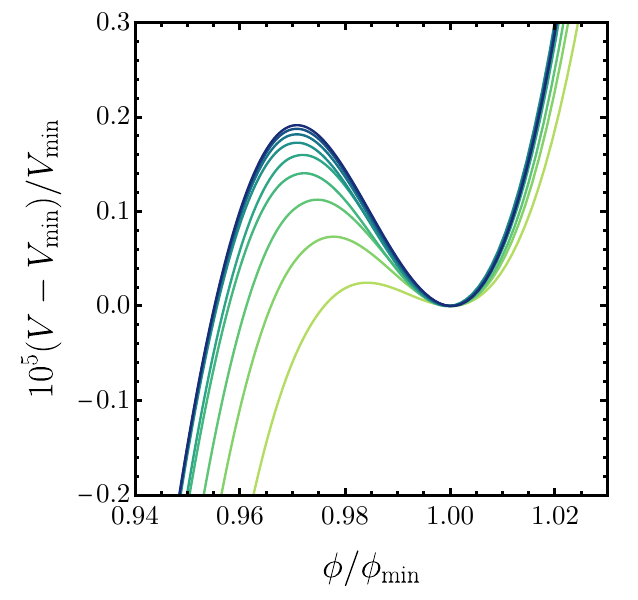} 
\caption{
\it Left panel: Potentials obtained by varying $\delta$ between $0.1$ (green) and $0.9$ (blue), for $\Delta N=2$. Right panel: Potentials obtained by varying $\Delta N$ between $1.1$ (green) and $2$ (blue), for $\delta=0.5$. We have set $\eta_{\rm USR}=4$ and $\eta_{\rm CR}=-1$ for both panels.
}
\label{fig:potentials_I}
\end{figure}

In this section we study the dependence of $\mathcal{I}$ and $\mathcal{N}$ on the sharpness of the SR-to-USR and the USR-to-CR transitions. To do this, we consider a simplification of a parametrization introduced in \cite{Taoso:2021uvl}, where a specific time-dependence (in terms of the number of e-folds, $N$) is imposed on the slow-roll parameter $\eta$:
\begin{equation}
\eta
=
\frac{1}{2}
\bigg[-\Delta\eta+(3+\Delta\eta)\tanh\bigg(\frac{N}{\delta }\bigg)-(3+2\Delta\eta)\tanh\bigg(\frac{N-\Delta N}{\delta }\bigg)
\bigg]\,,
\label{eq:etaparametrization}
\end{equation}
where we set $N = 0$ at the beginning of the USR phase.
Here $\delta$ is a positive number that controls the smoothness of the transitions in and out of USR (the smaller $\delta$, the sharper the transitions). The parameter $\Delta N$ controls, together with $\delta$, the duration of USR. Indeed this is approximately equal to $\Delta N-\delta$. For this reason, as we increase $\delta$, we also need to increase $\Delta N$ to keep the duration of USR (and therefore the heigth of the peak of the spectrum) more or less constant. The number $\Delta\eta$ characterizes the approximate values of $\eta$ in the USR and CR phases: $\eta_{\rm USR}=3+\Delta\eta$ and $\eta_{\rm CR}=-\Delta\eta$, respectively. This ensures that Wands duality \cite{Wands:1998yp} is satisfied\footnote{The validity of \eqref{eq:deltan_cr}  relies on $\dot\eta$ being negligible after the inflaton overcomes the local maximum of the potential (which is the definition of CR). This is ensured by Wands duality holding or, equivalently, the potential being accurately approximated by a parabola around the local maximum (see \cite{Ballesteros:2024pwn} for a recent discussion).} unless if $\delta$ is large enough  (in which case  $ \eta$ never reaches $\eta_{\rm USR}$). The shape of $\eta$ from \eqref{eq:etaparametrization} mimics (with smooth transitions) a piecewise constant function, with $\eta_{\rm USR}$ the value of $\eta$ in the USR phase and $\eta_{\rm CR}$ the one in the subsequent CR phase. Integrating $\eta$ in time, we recover $\epsilon$; the corresponding boundary condition is fixed by choosing a reasonably small value of $\epsilon$ in the SR phase. In particular, we choose $\epsilon = 10^{-3}$ at very early times. Integrating $\epsilon$, we recover $H$; in this case, for which the boundary condition is fixed by the amplitude of perturbations on very large scales, $\mathcal{P}_\zeta\simeq (H/M_P)^2/(8\pi^2\epsilon)\approx 2.2\times 10^{-9}$. The kind of potentials obtained by varying the parameters in this model are shown in Figure~\ref{fig:potentials_I}.

\begin{figure}[t!]
\centering
\includegraphics[width=0.49 \textwidth]{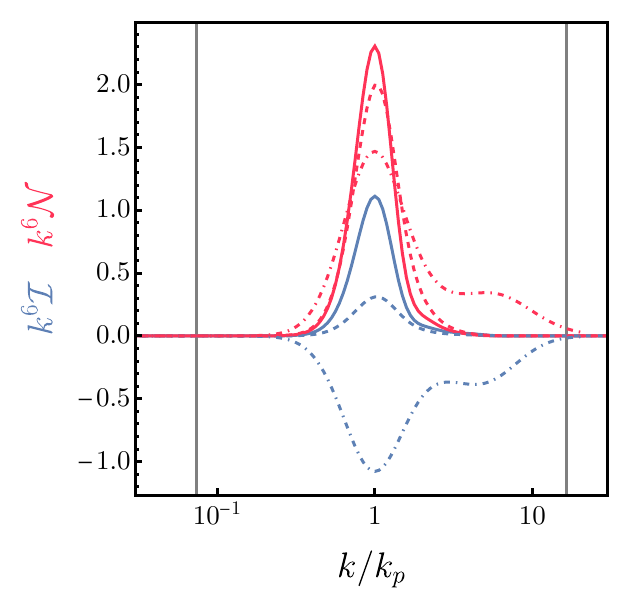} 
\includegraphics[width=0.49 \textwidth]{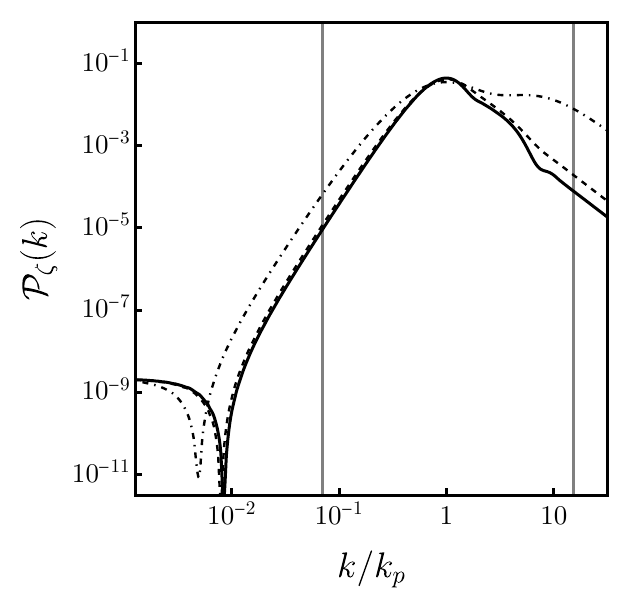} 
\caption{
\it Left panel: Contributions to the three-point function of $\zeta$ arising from the intrinsic non-Gaussianity of $\delta\phi$ (blue) and its nonlinear relation to $\zeta$ (red) rescaled by the factor $k^6$ for the sharp (solid), intermediate (dashed) and smooth (dot-dashed) examples in Section \ref{sec:num_model}. The vertical lines denote the approximate range of momenta probed by the lattice simulation presented in Section \ref{sec:lattice}. Right panel: Power spectra for the same examples as in the left panel. We denote the location of the peak in the spectra by $k_p$.
}
\label{fig:corrs_I}
\end{figure}

We choose three representative sets	 of parameters to illustrate our results (see Table~\ref{table:models}). The three examples are selected such that the power spectrum reaches $\mathcal{P}_\zeta(k_p)\simeq 10^{-2}$ at the peak,\footnote{This peak occurs for scales crossing the horizon at the USR-to-CR transition, $k_p \simeq a(\Delta N) H(\Delta N)$, as illustrated in Fig.~\ref{fig:corrs_I} on the right panel.} regardless of the sharpness of the transition. The consistency relation \eqref{eq:consistency} 
is illustrated for the three cases under consideration in Figure \ref{fig:consistency_I},
obtaining a good agreement. As we mentioned earlier, the fact that the consistency relation is satisfied is an indication that we are not neglecting relevant interactions when computing $\mathcal{I}$ including only the $c_0\,\varphi^3$ interaction.

\begin{table}[t]
\centering
\begin{tabular}{c || c   c c||} 
 & smooth & intermediate & sharp \\
 \hline
 $\delta $ & 0.90  & 0.33  & 0.22 \\
 $\Delta N$ & 2.19  & 1.72  & 1.63 \\
\end{tabular}
\caption{\it Parameters for the three illustrative models considered in this work. In all three cases, $\eta_{\rm USR}=4$ and $\eta_{\rm CR}=-1$ (in other words, $\Delta\eta=1$).}
\label{table:models}
\end{table}

We now move to the calculation of $\mathcal{I}$ and $\mathcal{N}$ using \eqref{eq:three_point} and \eqref{eq:calN}. 
As explained above, integrating \eqref{eq:etaparametrization} successively, we obtain $\epsilon$ and $H$. And differentiating \eqref{eq:etaparametrization}, we obtain $\epsilon_3$ and $\epsilon_4$, which allows to compute $c_0$, see Eq.\ \eqref{c0}. Also, having the slow-roll parameters, we can compute the time evolution of the Fourier modes $\varphi_k$ in the free theory through Eq.\ \eqref{ec:MS}.

{In order to compare the sizes of $\mathcal{I}$ and $\mathcal{N}$, we have to evaluate them at a certain momentum configuration. For illustrative purposes, we will consider the equilateral configuration of the bispectrum, where all the momenta coincide.}
We find that in this configuration the contributions $\mathcal{N}$ and $\mathcal{I}$ to the bispectrum are maximal on the scale associated with the peak of the power spectrum, see Figure \ref{fig:corrs_I}.
Therefore, a first assessment of the relative importance of the non-Gaussianity of $\delta\phi$ and the non-linear relation to $\zeta$ can be obtained with the ratio $\mathcal{I}(k_p)/\mathcal{N}(k_p)$.
If $\abs{\mathcal{I}(k_p)/\mathcal{N}(k_p)} \ll 1$, the non-Gaussianity of $\zeta$ is dominated by its non-linear relation to $\delta\phi$, making the non-Gaussianity of $\delta\phi$ unimportant. Therefore, in this case, one could expect Eq.\ \eqref{eq:pdf_r}, which assumes a Gaussian $\delta\phi$, to be a good approximation for the PDF of $\zeta$. However, if  $\mathcal{I}(k_p)/\mathcal{N}(k_p)$ is not small, $\mathcal{I}$ cannot be neglected and one may expect a non-linear transformation (such as Eq.\ \eqref{eq:deltan_cr}) of a Gaussian $\delta\phi$ to lead to a poor estimate of the PDF of $\zeta$. 

\begin{figure}
\centering
\includegraphics[width=0.49 \textwidth]{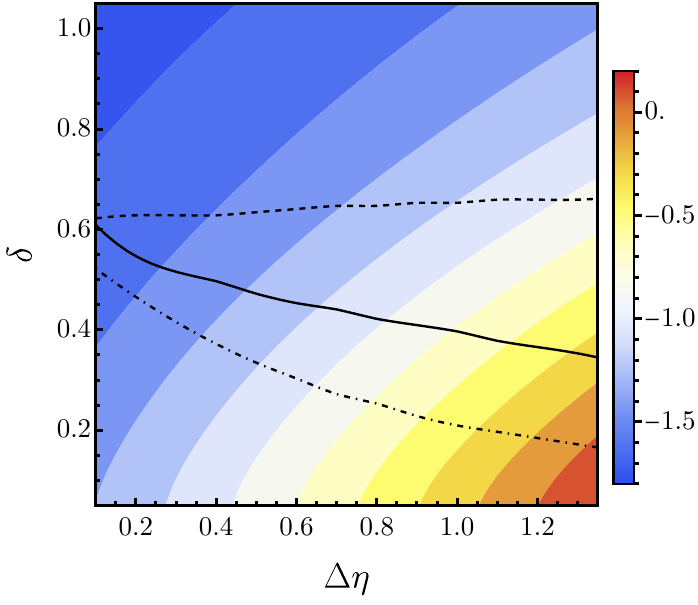} 
\includegraphics[width=0.49 \textwidth]{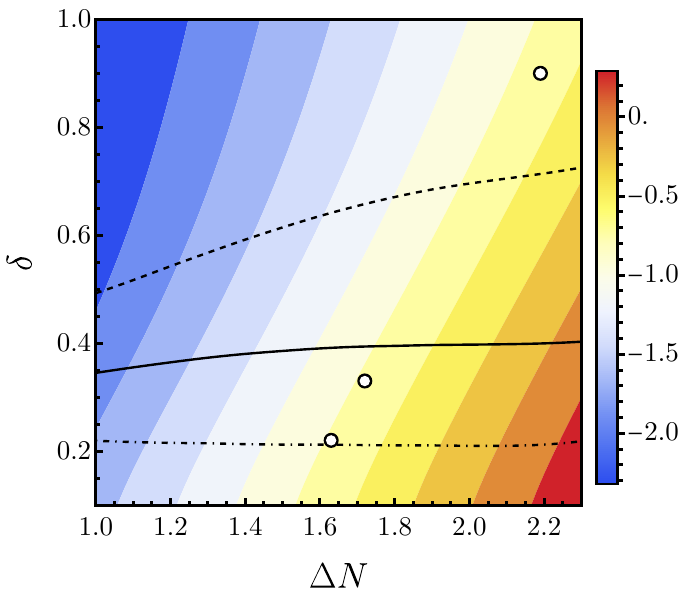} 
\caption{
\it Colored contours correspond to the peak value of $\log_{10}\mathcal{P}_\zeta$. Dashed, solid and dot-dashed black lines are the contours $\mathcal{I}/\mathcal{N}=-0.5,0,0.5$, respectively. Left panel shows the dependence of these two quantities on $\Delta \eta$ and $\delta$, fixing $\Delta N=1.7$. Right panel shows their dependence on $\Delta N$ and $\delta$, fixing $\Delta\eta = 1$. The three white spots correspond to the three cases described in Table \ref{table:models}, which indeed have a similar peak height in their power spectrum (Figure \ref{fig:corrs_I}, right panel). The assumption of Gaussian $\delta\phi$ is only valid sufficiently close to the solid black line.}
\label{fig:pars_I}
\end{figure}
 
The results obtained from this comparison are displayed in Figure \ref{fig:pars_I}.
We note that, in general, the contribution of $ \mathcal{I}$ to the bispectrum is not negligible with respect to that of $\mathcal{N}$. However, due to a change of sign of $\mathcal{I}$ around $\delta \simeq 0.4$, there is a region in the parameter space where the contribution of $\mathcal{I}$ to the bispectrum is highly suppressed (see Figure \ref{fig:pars_I}).
One can roughly estimate the value of $\delta$ for which $\mathcal{I}$ vanishes. 
First, one can notice that the factor $ {\rm Im}\big[ (\varphi_{k_p}(N) \varphi_{k_p}^*(N_e ) )^3\big]$ in Eq.\ \eqref{eq:three_point} peaks around $\Delta N$ and decreases for $N> \Delta N$ as $\varphi_{k_p}$ freezes to its asymptotic value.
In practice, this term acts as a window function that restricts the integral in \eqref{eq:three_point} to a narrow range of e-folds around the USR-to-CR transition, $N\simeq \Delta N$. 
Notice also that $c_0(N)$ crosses $0$ during this transition. Thus, for $N$ around $\Delta N$, one can minimize $\mathcal{I}$ by requesting $c_0'(\Delta N) = 0$. We find numerically that the most relevant contribution to $c_0'(\Delta N)$ around $N \simeq \Delta N$ is
\begin{equation}
\dfrac{\diff c_0}{\diff N} \simeq \dfrac{1}{3} \, \epsilon \, \eta \, \epsilon_3 \, \big( \epsilon + 2 \, \eta - \epsilon_3 - \epsilon_4 -3 \big)' \,,
\end{equation}
which vanishes if the derivative of the term in brackets vanishes. In the limit $\delta  \ll \Delta N$, this approximately corresponds to 
\begin{equation}
c_0'(N) \simeq(2 \Delta \eta +3)^2 [\delta   (2 \Delta \eta +3)-2] =0\implies \delta   =  \dfrac{2}{3+2\Delta \eta} =0.4\quad \text{for} \quad \Delta\eta=1\,,
\end{equation}
recovering the result represented in Figure \ref{fig:pars_I}. Notice that, as long as $\delta  \ll \Delta N$, this result is essentially independent of $\Delta N$.  

In \cite{Taoso:2021uvl}, it was argued that the non-Gaussianity of $\delta\phi$ is essentially negligible if the USR-to-CR transition is sufficiently smooth,  and the bispectrum of $\zeta$ is of the local form. It was found there that, as $\delta\to 0.44$, the contribution to the bispectrum of $\zeta$ due to the bispectrum of $\delta\phi$ vanishes, and the one coming from the field redefinition dominates. The conclusion they inferred was that, the smoother the transition, the more subdominant the intrinsic non-Gaussianity will become. However, we have seen that $\mathcal{I}$ vanishes for $\delta\approx 0.4$ because it crosses zero, not because it vanishes asymptotically. Indeed, if we keep increasing the value of $\delta$, $\mathcal{I}$ (the intrinsic non-Gaussianity) becomes sizable again, only with a different sign.\footnote{The analysis in \cite{Taoso:2021uvl} was done in the $\zeta$-gauge, using Maldacena's field redefinition $\mathcal{R}\to\mathcal{R}_n+f(\mathcal{R}_n)$ (see Eqs.\ (A4-A6) therein). Their $\mathcal{R}_n$ corresponds with our $\varphi$ in the $\delta\phi$-gauge.}

In the next section, we will provide a more refined assessment of the relative contribution of $\mathcal{I}$ and $\mathcal{N}$ to the non-Gaussianity of $\zeta$.

\section{Lattice simulation}
\label{sec:lattice}

In the last section it has been shown that the effect of the intrinsic non-Gaussianities of $\zeta$ --understood to be those coming from the interactions of $\delta\phi$-- on its bispectrum can be at least as important as the contribution associated with the non-linearities between $\zeta$ and $\delta\phi$, depending on the parameters of the model. Therefore, a correct description of the PDF of $\zeta$ (and particularly its tail) may require including both sources of non-Gaussianities. However, there is no straightforward way to incorporate non-Gaussian corrections into 
the PDF starting from a Gaussian distribution.\footnote{This is what the Gram-Charlier (or Edgeworth) series tries to do, but it does not converge in the present case. This series can even lead to non-positive results for the PDF upon truncation at certain orders, which clearly limits its usefulness to expand the PDF (see Appendix \ref{app:gram-charlier} for more details).}

In this section we present a numerical method to generate a non-Gaussian field satisfying specific input perturbative statistics that we can choose at will, starting from a stochastic variable following a Gaussian distribution. 
In particular, we will construct a stochastic variable $\tilde \zeta$ --where from now on $\, \tilde{\,\,}\,$ will be carried by stochastic variables-- in such a way that its statistical moments  coincide with the $n$-point correlation functions of the variable $\hat{\zeta}$ obtained from the in-in formalism, i.e.\
\begin{equation}
	\langle \tilde \zeta_{\boldsymbol{k}_1}\tilde \zeta_{\boldsymbol{k}_2}\cdots\tilde \zeta_{\boldsymbol{k}_n} \rangle_c  \, = \,   \langle \hat \zeta_{\boldsymbol{k}_1} \hat \zeta_{\boldsymbol{k}_2}\cdots\hat \zeta_{\boldsymbol{k}_n} \rangle_c \,.
\end{equation}
The brackets $\langle \cdots \rangle$ refer to ensemble averages or correlation functions when applied to stochastic variables or quantum operators, respectively. 
We recall that the subscript $_c$ stands for the connected part of the averages and correlations, and it always comes accompanied by $\delta(\bm{k}_1 + \cdots + \bm{k}_n)$ that guarantees the momentum conservation. Imposing that the connected parts of the $n$-point ensemble averages and quantum correlation functions coincide, it is guaranteed that their disconnected parts also coincide, because the disconnected parts are obtained by the product of connected parts of lower orders \cite{Ballesteros:2024qqx}.

Once we have ensured that $\tilde{\zeta}$ has the correct statistic, we can sample $\tilde{\zeta}$ to obtain its non-Gaussian probability distribution, thus completing the calculation of the PDF.

Since the relation between the curvature fluctuation $\zeta$ and the inflaton fluctuation $\delta\phi$ in the $\delta\phi$-gauge is exactly described by Eq.~(\ref{eq:deltan_cr}), we can describe $\tilde{\zeta}$ implicitly by means of a variable $\tilde{\varphi}$ that satisfies the correct statistics,	
\begin{equation}
	\langle \tilde \varphi_{\boldsymbol{k}_1}\tilde \varphi_{\boldsymbol{k}_2}\cdots\tilde \varphi_{\boldsymbol{k}_n} \rangle_c  \, = \,   \langle \hat \varphi_{\boldsymbol{k}_1} \hat \varphi_{\boldsymbol{k}_2}\cdots\hat \varphi_{\boldsymbol{k}_n} \rangle_c \,,
	\label{eq:correlationslattice}
\end{equation} 
as discussed above.
We stress that the choice of $\varphi$, rather than $\delta\phi$, is simply for convenience; but it is worth noting that $\varphi$ still describes the scalar fluctuations in the $\delta\phi$-gauge, Eq.~\eqref{ec:varphi}. 

Let us emphasize that our procedure is not equivalent to performing a dynamical lattice simulation of the time evolution of the scalar fluctuations during inflation, as in \cite{Caravano:2024moy}.
Instead, we calculate the statistical properties of the scalar fluctuation $\zeta$ using the in-in formalism, and simulate its probability distribution {\it at the end of inflation}, where the modes of $\zeta$ with sufficiently long wavelength  will be frozen at all orders in perturbations \cite{Maldacena:2002vr,Lyth:2004gb,Weinberg:2003sw} (see also \cite{Pimentel:2012tw,Assassi:2012et,Senatore:2012ya,Green:2024fsz} for a discussion of this statement at the quantum level).\footnote{{Although at the end of inflation there are still modes inside the horizon, and therefore not all the modes are frozen, since we are interested in the physics around the USR phase, it is legitimate to assume that in those times all the modes of interest will be frozen.}}
This allows us to incorporate the quantum dynamical effects arising from the in-in formalism.
The price to pay is that, following our procedure, the probability distribution can only be found perturbatively, because the correlation functions are obtained by means of a loop (or Planck mass) expansion using the in-in formalism.

Although our procedure to obtain the PDF of $\tilde{\varphi}$ can be applied at any time during inflation, the change of gauge we use subsequently to get the PDF of $\tilde{\zeta}$ works only when the coarse-grained variable $\zeta$ is frozen, i.e.\ when all the Fourier modes of interest have become constant. In practice, for the kind of models we are interested in, this is equivalent to applying the change of gauge at the end of inflation, provided that the enhancement of $\zeta$ occurs for modes that become super-Hubble early enough.

\subsection{Non-Gaussian field generation}

To construct the stochastic variable $\tilde{\varphi}$, we will make use of an auxiliary Gaussian stochastic variable of unit variance $\tilde{\psi}_{\bm k}$, whose statistics is fully determined by the two-point correlation function:
\begin{equation}
	\langle \tilde{\psi}_{\bm k} \tilde{\psi}_{\bm p}\rangle
	=
	(2\pi)^3\delta^{(3)}({\bm k}+{\bm p})\,.
\end{equation}
Since $\tilde{\psi}_{\bm k}$ is Gaussian, any $n$-point correlator can be obtained by using Wick's theorem. We decompose $\tilde{\varphi}$ as follows:
\begin{equation}
	\label{eq:phiAnsatz}
	\tilde{\varphi}(\bm{x}) = \int\frac{\diff^3k}{(2\pi)^3} e^{i \bm{k} \bm{x}} \tilde{\varphi}_{\bm k}\,, \quad \textrm{with} \quad
	\tilde{\varphi}_{\bm k}=\sum_n\tilde{\varphi}^{(n)}_{\bm k}\,,
\end{equation}
where
\begin{equation}
	\tilde{\varphi}^{(n)}_{\bm k}
	\equiv
	\frac{1}{n!}
	\int\frac{\diff^3k_1}{(2\pi)^3}\cdots \frac{\diff^3k_{n-1}}{(2\pi)^3}\mathcal{G}_n({\bm k},-{\bm k}_1,\cdots, -{\bm k}_{n-1}) \frac{\tilde{\psi}_{{\bm k}_1}}{|\varphi_{k_1}|}
	\cdots
	\frac{\tilde{\psi}_{{\bm k}_{n-1}}}{|\varphi_{k_{n-1}}|}
	(2\pi)^3
	\delta^{(3)}(-{\bm k}+{\bm k}_1+\cdots +{\bm k}_{n-1}) \,.
	\label{eq:phi_stoch_general}
\end{equation}
We emphasize that we are describing the variable $\tilde{\varphi}$ at the end of inflation, $t_e$, so we omit the time dependence of all the objects that compose it assuming that they are evaluated at that time.
Due to the freedom we have in choosing the functions $\mathcal{G}_n$, as long as they are totally symmetric under momenta permutations, we will be able to impose that $\tilde{\varphi}$ satisfies the appropriate statistics, as we previously discussed. 
The functions $\varphi_{k_i}$ are the modes, solution to the free equation of motion, which we find in the operator $\hat{\varphi}$ in Eq.~(\ref{eq:varphi int pic}). They are evaluated at the end of inflation and therefore we omit their time dependence.
For convenience, we also define the connected part of the $n$-point correlation of $\hat{\varphi}$ at the end of inflation as:
\begin{align} \label{eq:ConnectedCorrelation}
	\langle\hat{\varphi}_{\bm k_1} \cdots \hat{\varphi}_{\bm k_n}\rangle_c \equiv
	\mathcal{I}_n({\bm k}_1,\cdots,{\bm k}_n)
	(2\pi)^3\delta^{3}({\bm k_1}+\cdots+{\bm k_n})\,.
\end{align}
We can thus identify $\mathcal{I}_3=\mathcal{I}$ as the intrinsic contribution to the bispectrum in Eq.\ (\ref{eq:three_point}).

We will assume perturbation theory to be valid. This means that the correlators $\mathcal{I}_n$, Eq.\ \eqref{eq:ConnectedCorrelation}, obtained through the in-in formalism will be tree-level dominated; with loop level corrections, generated by further insertions of the interaction Hamiltonian, see Section \ref{sec:In-in}, being negligible. We will discuss the validity of this assumption for our examples later on. 

Naively, the parameter controlling the validity of perturbation theory should be the ratio $H / M_P$. Thus, we can estimate the order of magnitude of $\mathcal{I}_n$ just by counting Planck masses. At $n$-th order, the interaction vertex in the $\delta\phi$-gauge is $\mathcal{L}_n\sim M_P^{2-n}$. Therefore, $\mathcal{I}_n \sim \left( H / M_P \right) ^{2n-2}$, where the additional $M_P^{-n}$ factor comes from each of the $n$-external fields $\varphi$ coupled to the interaction vertex generated by $\mathcal{L}_n$, and where we have introduced $H$ because it is the only physical scale that can accompany $M_P$. 
As we will see below, $\mathcal{G}_n \sim I_n$ and finally $\tilde{\varphi}^{(n)}_{\bm k} \sim \mathcal{G}_n |\varphi_k|^{-(n-1)} \sim \left(H / M_P \right) ^{n-1}$. Therefore, the expansion through which we define $\tilde{\varphi}$, Eq.~\eqref{eq:phiAnsatz}, corresponds to a perturbative expansion in powers of $H/M_P$. 

Let us now deduce the functions $\mathcal{G}_n$ imposing Eq.\ \eqref{eq:correlationslattice} order by order. Although the first correlation to be analyzed is the one-point function, associated to the zero mode of $\tilde{\varphi}$ and therefore indistinguishable from the background, we will impose in each simulation that it is absorbed by the background.\footnote{The one-point correlation is a special case, as its leading order is the tadpole, since there is no tree-level contribution. This means that, in practice, this contribution corresponds to a negligible shift on the background, which further justifies that we omit its precise treatment.}

Analyzing the connected two-point correlation at leading order, we obtain:
\begin{equation}
	\langle\tilde{\varphi}_{\bm k}\tilde{\varphi}_{\bm p}\rangle_c = \langle \tilde{\varphi}^{(2)}_{\bm k} \tilde{\varphi}^{(2)}_{\bm p}\rangle_c = \frac{1}{4}
	\frac{\mathcal{G}_2({\bm k},{\bm p})^2} {|\varphi_{k}||\varphi_{p}|} (2\pi)^3\delta^{(3)}({\bm k}+{\bm p})\,.
\end{equation}
This connected average also receives corrections from $\tilde{\varphi}^{(n)}_{\bm k}$ with $n>2$, but they will be suppressed by powers of $H / M_P$. Imposing the condition of Eq.\ \eqref{eq:correlationslattice}, and taking into account that $\mathcal{I}_2(\bm{k},\bm{p}) = |\varphi_{k}||\varphi_{p}|$, we get 
\begin{equation}\label{ec:g2}
	\mathcal{G}_2({\bm k},{\bm p}) = 2|\varphi_{k}||\varphi_{p}|\,, \quad \textrm{and thus} \quad \tilde{\varphi}^{(2)}_{\bm k} = |\varphi_{k}|\, \tilde{\psi}_{\bm{k}}\,.
\end{equation}
Repeating the same procedure with the three-point correlation gives
\begin{equation}
	\mathcal{G}_3({\bm k},{\bm p},{\bm q}) 	= \mathcal{I}_3({\bm k},{\bm p},{\bm q})\,,
\end{equation}
while for the four-point correlation
\begin{equation}
	\mathcal{G}_4({\bm k},{\bm p},{\bm q},{\bm \ell}) =	\mathcal{I}_4({\bm k},{\bm p} ,{\bm q},{\bm \ell}) - \frac{4}{9} 	\bigg[ 	\frac{ \mathcal{I}_3({\bm k},-{\bm p}-{\bm k},{\bm p}) \mathcal{I}_3({\bm q},-{\bm \ell}-{\bm q},{\bm \ell})}{ |\varphi_{|{\bm p}+{\bm k}|}|^2 } +	({\bm p}\rightarrow {\bm q},{\bm \ell}) 	\bigg]\,.
\end{equation}
In this last case, $\mathcal{G}_4$ receives a contribution not only from $\mathcal{I}_4$ but also from two insertions of $\mathcal{I}_3$. This is to be expected, since $\mathcal{I}_4$ at tree-level is composed of a contact diagram and diagrams in which a virtual scalar field propagates in an $s$, $t$ and $u$ channel.

The procedure can be extended in a straightforward manner to arbitrary orders, increasing the precision with which we describe the statistics of the variable $\tilde{\varphi}$. However, to illustrate how the corrections to the Gaussian case (described exclusively by $\mathcal{G}_2$) modify the PDF of $\tilde{\varphi}$ it is sufficient to just include $\mathcal{G}_3$, as we will explore in the next section. In general, the effect on the PDF of including increasingly higher-order corrections $\mathcal{G}_n$ will be a modification in the tail of the distribution, increasingly further away from the origin. Therefore, truncating the expansion up to a certain order will allow us to accurately describe the PDF up to a certain value of the variable $\tilde{\varphi}$, provided that perturbation theory works. 

We stress that, as explained above, by imposing \eqref{eq:correlationslattice}, the disconnected contributions to the $n$-point functions also coincide between averages and correlations, 
\begin{equation}
	\langle \tilde \varphi_{\boldsymbol{k}_1}\tilde \varphi_{\boldsymbol{k}_2}\cdots\tilde \varphi_{\boldsymbol{k}_n} \rangle  \, = \,   \langle \hat \varphi_{\boldsymbol{k}_1} \hat \varphi_{\boldsymbol{k}_2}\cdots\hat \varphi_{\boldsymbol{k}_n} \rangle \,,
\end{equation}
which can be explicitly verified using the definition of $\tilde{\varphi}$ in Eq.\ \eqref{eq:phiAnsatz} and the functions $\mathcal{G}_n$ described above.

\subsection{{Validity of perturbation theory}} \label{sec:valid}

Let us analyze in more detail the validity of the perturbative expansion through which we describe $\tilde{\varphi}$, Eq.\ \eqref{eq:phiAnsatz}. Although, as we have seen, the naive parameter controlling the expansion is $H/M_P$, the details of the interactions may mean that even though $H/M_P\ll 1$, in fact perturbation theory breaks. To guarantee the validity of perturbation theory, it has to be ensured that the different averages are dominated by the leading order. For example, in the two-point case,
\begin{equation}\label{ec:valid_pert}
	\langle\tilde{\varphi}_{\bm k}\tilde{\varphi}_{\bm p}\rangle_c = \langle \tilde{\varphi}^{(2)}_{\bm k} \tilde{\varphi}^{(2)}_{\bm p}\rangle_c + \langle \tilde{\varphi}^{(3)}_{\bm k} \tilde{\varphi}^{(3)}_{\bm p}\rangle_c + \langle \tilde{\varphi}^{(2)}_{\bm k} \tilde{\varphi}^{(4)}_{\bm p}\rangle_c + \cdots\,,
\end{equation}
it has to be satisfied that the latter two contributions are much smaller than the first one. Diagrammatically, these last two contributions are identified with the calculation of the two-point correlation at one loop. Therefore, we can state that our perturbative expansion will be valid, in general, when the loop-level corrections to the $n$-point correlation functions are suppressed with respect to the tree-level contribution. Including loop-level corrections would be possible by modifying $\mathcal{G}_n$, but it is beyond the scope of this paper.

\begin{figure}
	\centering
	\includegraphics[width=0.49 \textwidth]{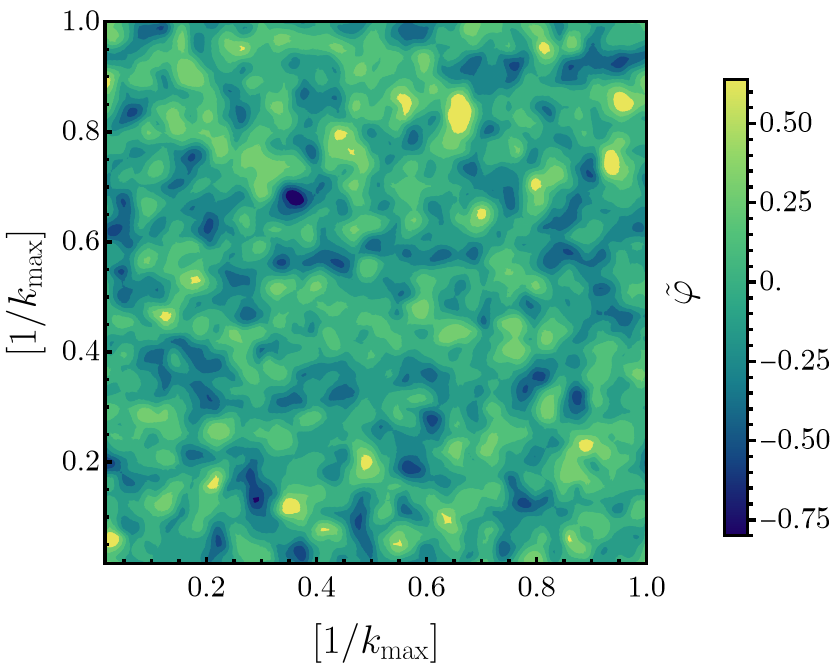} 
	\includegraphics[width=0.475 \textwidth]{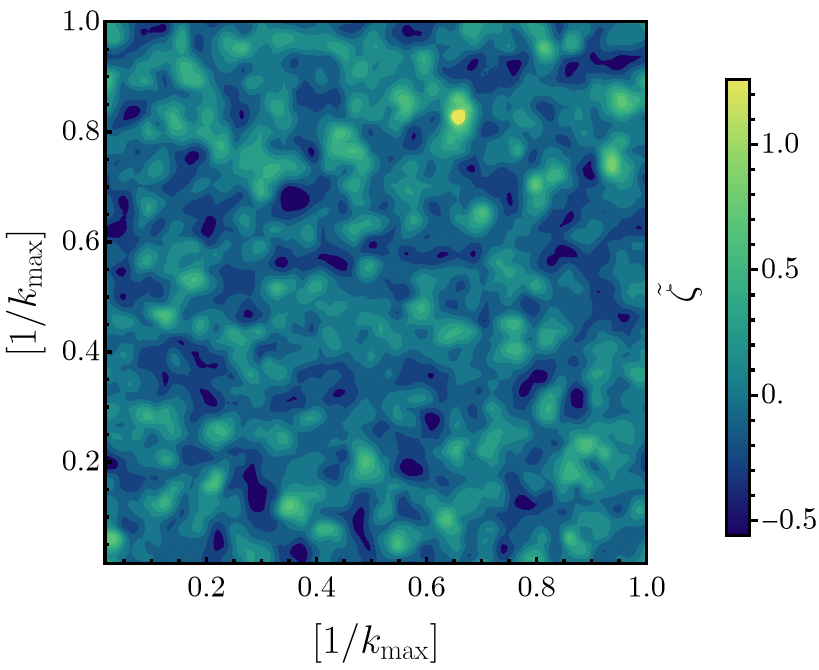} 
	\caption{
		\it Left panel: Two-dimensional slice of one realization of the intermediate transition example in Section \ref{sec:num_model} for the Gaussian contribution to $\tilde \varphi$ (i.e. $\tilde \varphi^{(2)}$). Right panel: Same realization as in the left panel for the full variable $\tilde \zeta$. Due to the strong non-Gaussian tail, hot spots are pushed to larger values.
	}
	\label{fig:ng_field}
\end{figure}

\subsection{Numerical results}

In this section, for concreteness and to illustrate the method that we just described, we analyze the effect of the bispectrum of $\delta\phi$ on the non-Gaussian tail of the PDF of $\zeta$. Using the relations obtained in the previous section, we have
\begin{equation}\label{ec:ref1}
	\tilde{\varphi}^{(3)}_{\bm k}
	=
	\frac{1}{3!}\int\frac{\diff^3p}{(2\pi)^3}
	\mathcal{I}({\bm k},{\bm k}-{\bm p},-{\bm p})
	\frac{\tilde{\psi}_{{\bm k}-{\bm p}}}{|\varphi_{|{\bm k}-{\bm p}|}|}
	\frac{\tilde \psi_{\bm p}}{|\varphi_p|}\,.
\end{equation}
Let us rewrite this expression in a way that is numerically more tractable. We define the function
\begin{equation} \label{redefe}
	\lambda_k(t)
	\equiv
	\varphi_k(t_e)\varphi_k^*(t)
\end{equation}
in terms of the free modes $\varphi_k(t)$, see Eq.\ \eqref{eq:varphi int pic}. Using Eq.\ \eqref{eq:three_point}, we find, at the end of inflation:
\begin{equation}
	\tilde{\varphi}^{(3)}_{\bm k} = -i M_P^2\,\int\frac{\diff^3p}{(2\pi)^3} \int_{-\infty}^{t_e} \diff t \; a(t)^3 c_0(t) H(t)^2 \Big[ \lambda_p(t) \lambda_{|{\bm k}-{\bm p}|}(t)
	\lambda_k(t) -{\rm h.c.} \Big] \frac{\tilde{\psi}_{{\bm k}-{\bm p}}}{|\varphi_{|{\bm k}-{\bm p}|}|} \frac{\tilde \psi_{\bm p}}{|\varphi_p|}\,,
\end{equation}
where, again, the modes $\varphi_k$, without an explicit time dependence, are evaluated at the end of inflation, $t_e$.
Defining
\begin{equation}
	\tilde{\Psi}^+_{\bm k}(t) \equiv \frac{\lambda_k(t)}{|\varphi_k|}\tilde \psi_{\bm k} \quad {\rm and} \quad \tilde{\Psi}^-_{\bm k}(t) \equiv \frac{\lambda_k^*(t)}{|\varphi_k|}\tilde \psi_{\bm k}
\end{equation}
in terms of the auxiliary variable $\tilde{\psi}_{\bm{k}}$, we can rewrite $\tilde{\varphi}^{(3)}_{\bm k}$ as:
\begin{equation}
	\tilde{\varphi}^{(3)}_{\bm k} = -i M_P^2\, \int_{-\infty}^{t_e} \diff t \; a(t)^3 c_0(t) H(t)^2
	\bigg\{ \lambda_k(t) \big[\tilde{\Psi}^+(t)^2\big]_{\bm k} - \lambda^*_k(t)
	\big[\tilde{\Psi}^-(t)^2\big]_{\bm k} \bigg\}\,, \label{eq:final_cubic}
\end{equation}
where
\begin{equation}\label{ec:aux_lambda}
	\big[\tilde{\Psi}^\pm(t)^2\big]_{\bm k} = \int\frac{\diff^3p}{(2\pi)^3} \tilde{\Psi}^\pm_{\bm p}(t) \tilde{\Psi}^\pm_{{\bm k}-{\bm p}}(t) 
\end{equation}
denotes the Fourier transform of $\tilde{\Psi}^\pm(t)^2$. 
We emphasize that writing the momentum integral of $\tilde{\varphi}^{(3)}_{\bm k}$ as a product in coordinate space makes the numerical calculation much more efficient.\footnote{Writing $\tilde{\varphi}^{(n)}_{\bm k}$ as a product in coordinate space  can be done at higher orders, $n>3$, only with the contributions associated with the contact diagram. In those diagrams where there is a certain internal propagation, we will not be able to write the convolution and will be required to perform the momentum integral explicitly, increasing the computational cost.}

\begin{figure}
	\centering
	\includegraphics[width=\textwidth]{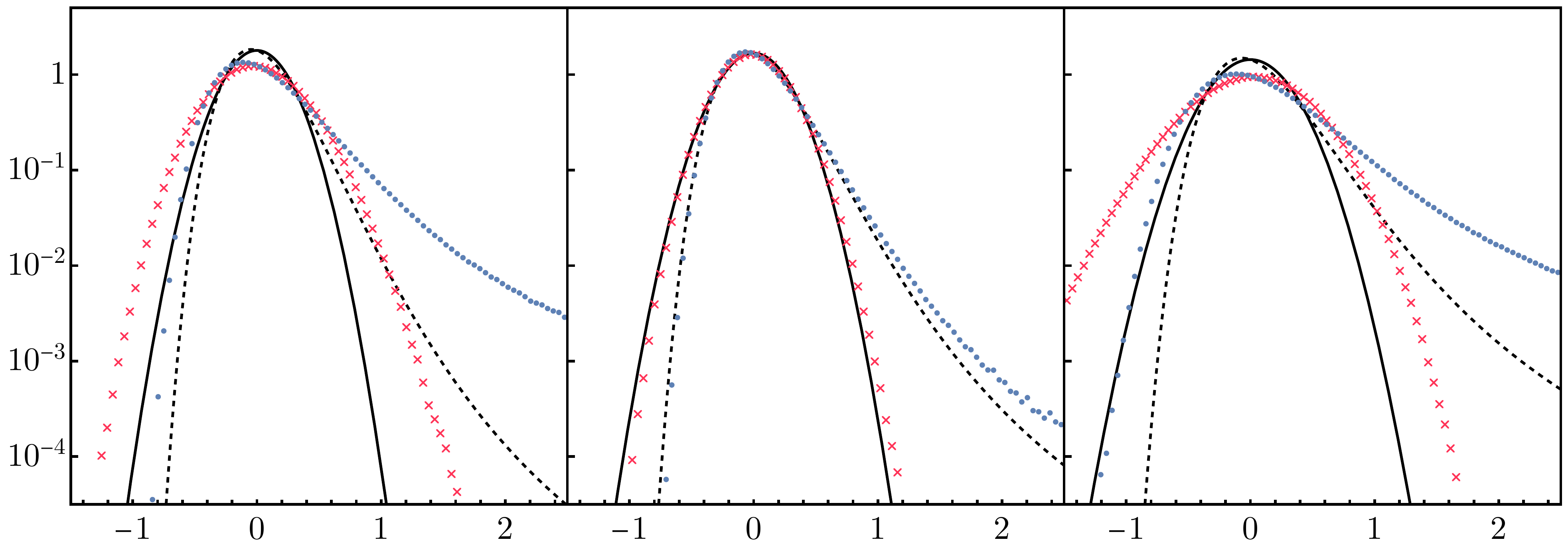}
	\caption{
		\it PDFs for the curvature perturbation $\tilde \zeta$ and the variable $\tilde \varphi$ for the three examples in Section \ref{sec:num_model}. The panels correspond, from left to right, to a sharp, intermediate, and smooth transition in and out of USR. The solid and dashed black lines denote the Gaussian contribution to $\tilde \varphi$ (i.e. $\tilde \varphi^{(2)}$) and the corresponding distribution for~$\tilde \zeta$, respectively. The red crosses denote the reconstructed non-Gaussian PDF for $\tilde \varphi$ (i.e. $\tilde \varphi^{(2)}+\tilde \varphi^{(3)}$), and the blue dots the corresponding PDF for $\tilde \zeta$ after using Eq.\,(\ref{eq:deltan_cr}).
	}
	\label{fig:pdf_I}
\end{figure}

Once the stochastic variable $\tilde{\varphi}$ has been characterized through the auxiliary variable $\tilde{\psi}_{\bm{k}}$, we are in the position to analyze its statistics and to obtain its PDF by performing a sampling. 
As we discussed, given the PDF $\tilde{\varphi}$, obtaining the PDF of $\tilde{\zeta}$ is straightforward by their non-linear relation, described in Eq.\ \eqref{eq:deltan_cr}.

In our simulation, we consider a lattice with $N^3=256^3$ points. In real space, it is defined within a certain volume $L^3$ and has a resolution of $l^3$. This naturally establishes cutoffs in momentum space, $k_{\rm min} = 2\pi / L$ and $k_{\rm max} = 2\pi / l$. {As we discussed at the end of Section \ref{sec:delta N formula}, since we are interested in PBH formation, we must choose these cutoffs such that they resolve the physics involved in the USR phase, as shown in Figure \ref{fig:corrs_I}}.\footnote{For practical purposes, one can take $l=1, L=N$, such that $k_{\rm max} = 2\pi $ and $k_{\rm min} = 2\pi / N$, and normalize the momenta when solving the equation {of motion of $\varphi_k$} such that the peak of the power spectrum roughly coincides with $2\pi/\sqrt{N}$. Also, in our numerical calculation, we stick to the DFT conventions in the FFT module of Numpy. In particular, we define the inverse DFT with a prefactor of $1/N$, consistent with the factor of $1/(2\pi)^3$ in the inverse Fourier transform used for the in-in calculations.} 
The main steps of our simulation are summarized below:
\begin{enumerate}
	\item Generate a random (real) Gaussian field with unit variance and Fourier-transform it to obtain $\tilde\psi_{\bm k}$. For each node in the lattice in momentum space, use Eq.\ \eqref{ec:g2} to compute $\tilde{\varphi}_{\bm k}^{(2)}$, and inverse Fourier-transform to obtain $\tilde{\varphi}^{(2)}(\bm{x})$.
	\item Start  with the variable $\tilde\psi_{\bm k}$ constructed in the previous step. 
	For each node in the lattice in momentum space, use Eq.\ \eqref{eq:final_cubic} to compute $\tilde{\varphi}_{\bm k}^{(3)}$.\footnote{This implies computing one lattice for each time step in the integrand, and then integrating numerically over the samples of the integrand; in our case, we found a time step of $0.15$ e-folds to be enough to accurately compute the integral using Simpson's method.} This requires computing the convolution in \eqref{ec:aux_lambda}, which is more efficiently done using the convolution theorem: transform $\tilde{\Psi}^+_{\bm k}(t)$ to real space, square it, and transform the result back to Fourier space. Finally, inverse Fourier transform $\tilde{\varphi}_{\bm k}^{(3)}$ to obtain $\tilde{\varphi}^{(3)}(\bm{x})$.
	\item Adding up $\tilde{\varphi}^{(2)}(\bm{x})+\tilde{\varphi}^{(3)}(\bm{x})$, we get a random field with the right power spectrum and bispectrum (up to $H/M_P$ suppressed corrections, as discussed above), whose PDF can be estimated numerically constructing a histogram. The PDF for $\tilde\zeta$ can then be obtained using the non-linear relation between $\zeta$ and $\var\phi$.
\end{enumerate}

The resulting PDF for $\tilde\zeta$ --composed by Fourier modes becoming super-Hubble around the USR phase, as defined by the lattice cutoffs-- is shown in Figure \ref{fig:pdf_I} for the examples of Table \ref{table:models}.

Neglecting the non-Gaussianity of the inflaton perturbation $\delta\phi$ to describe the PDF of $\tilde{\zeta}$, as usually done in stochastic inflation or the $\delta N$ formalism, i.e.\,by using only Eq.\,(\ref{eq:pdf_r}), may not be a good approximation, depending on the duration of the USR phase and the abruptness of the transitions between SR and USR and USR and CR. In fact, the Gaussian approximation only works in a small region of parameter space in the model we have considered (see also Figure \ref{fig:pars_I} and the corresponding discussion). We also recall that the region in which the intrinsic contribution to the bispectrum becomes negligible is related to a change of sign in the function $\mathcal{I}$, so that the validity of Eq.\,(\ref{eq:pdf_r}), when it happens, is due to an accidental cancellation. {Incidentally, note that different momentum configurations of $\mathcal{I}$ enter the integral in \eqref{ec:ref1} to compute $\tilde{\varphi}^{(3)}_{\bm k}$, aside from the equilateral limit studied in Sec.~\ref{sec:correlators}. As already emphasized, the reason to choose this limit in Sec.~\ref{sec:correlators} was to obtain some simple intuition on the hierarchy between the different sources of non-Gaussianity before moving into the main calculation in the present section.}

The examples we have labeled ``smooth'' and ``sharp'' (referring to the abruptness of the transitions) display very large changes in the PDF of $\zeta$ once the bispectrum of $\delta\phi$ is taken into account. However, perturbation theory breaks for these examples in the sense discussed in Section \ref{sec:valid}. Indeed, we have checked numerically that the variance of $\tilde{\varphi}^{(3)}$ is comparable to that of $\tilde{\varphi}^{(2)}$.  In these cases, our method cannot be applied to derive the correct PDF and can only be used as an illustration that significant deviations from the purely Gaussian approximation can be expected. Truly {non}-perturbative methods would be required to deal with these examples and other choices of parameters for which perturbation theory breaks. Indeed, in such cases the outcome cannot be improved by including the effects of a finite number of higher-order correlators of $\delta\phi$. 

The example we have labeled ``intermediate'' is interesting because it shows an appreciable difference between including the bispectrum of $\delta\phi$ or not, in a case in which the variances of $\tilde{\varphi}^{(2)}$ and $\tilde{\varphi}^{(3)}$ are, respectively,  $\mathcal{O}(10^{-2})$ and $\mathcal{O}(10^{-3})$. We can get an idea of the importance of the effect by computing the integral of the PDF (with and without the inclusion of the bispectrum of $\delta\phi$) from some value of $\zeta$ up to the largest $\zeta$ for which we have data. This integral can be regarded as a first proxy to the calculation of the abundance of PBH, which  is highly sensitive to the tail of the PDF. While an actual evaluation of the abundance is a more complex task, this integral is sufficient to illustrate in simple terms the relevance of including the bispectrum. For instance, integrating from $\zeta = 0.5$ we get that the ratio of the integrals is $\sim 1.5$ and if we integrate from $\zeta = 1$, we obtain $\sim 2$ for the same ratio; with the ratio changing linearly as a function of the lower limit of integration.

\phantomsection
\section{Summary and discussion}

The assumption of Gaussian inflaton perturbations commonly used to derive the non-Gaussian PDF of $\zeta$  from its non-linear relation to $\delta\phi$ can be insufficient. This can happen for large enough $\zeta$ (which can be relevant for phenomena such as PBH formation) provided that self interactions of $\delta\phi$ are non-negligible. An interesting scenario in this context is that of a transient USR phase that enhances curvature fluctuations at specific comoving scales.

 It is therefore important to develop methods that allow to compute the tail of the PDF including the effect of the {\it intrinsic} non-Gaussianities of $\delta\phi$, i.e.\ its self interactions. In this work we have approached this problem developing a method that allows to include these non-Gaussianities perturbatively by simulating them on a lattice. By construction, the method can be applied as long as perturbation theory does not break, but it can nonetheless provide interesting insights even when its regime of applicability starts to fail. 

We have illustrated the method parametrizing the second slow-roll parameter, $\eta$, which allows to fix the sharpness of the transitions between the phases of inflation (SR, USR and CR) with a number parameter, $\delta$, which controls approximately the duration of said transitions. We determined the tree-level bispectrum of $\delta\phi$ using the in-in formalism and validated our choice of interaction terms checking Maldacena's consistency relation. In practice, we isolated the terms in the cubic interaction Hamiltonian (in the $\delta\phi$-gauge) providing the most relevant contribution to the bispectrum of $\delta\phi$. These turn out to be the self interactions of the form $\delta\phi^3$, coming from the third derivative of the potential of the inflaton. As a first estimate of the relevance of intrinsic non-Gaussianities, we compare the contribution to the bispectrum of $\zeta$ arising from the bispectrum of $\delta\phi$ and the contribution due to a (perturbative) quadratic relation between $\delta\phi$ and $\zeta$, which is non-zero even for a Gaussian $\zeta$. We find that the former contribution is only negligible with respect to the latter for a small range of values of $\delta$ (when the bispectrum of $\delta\phi$ crosses zero as it changes sign). This result already suggests that, in general, one should not neglect intrinsic non-Gaussianities when providing an input for a non-linear relation between $\delta\phi$ and $\zeta$. 

To further account for this non-linear relation up to all non-linear orders (in the relation between $\zeta$ and $\delta\phi$) we used the $\delta N$ formalism (which is nothing but a change of gauge) and applied our numerical lattice procedure. We began by generating configurations of values of $\delta\phi$ (in position space) featuring the tree-level power spectrum and bispectrum, which is enough to illustrate the effects of the non-Gaussianities on the PDF of $\zeta$. We then sample this random field and use its PDF as an input for its non-linear relation to $\zeta$ provided by the $\delta N$ formalism. This gives the change in the PDF of the non-linear $\zeta$ obtained by accounting for the non-zero bispectrum of $\delta\phi$. As anticipated by our first estimate --see the previous paragraph--, the deviations from the naive result assuming Gaussian $\delta\phi$ can be sizable, except for a narrow range of values of $\delta$. 

Large variations of the inferred, {\it intrinsically} modified, PDF occur when perturbation theory starts to break, which underlines the need of developing fully non-perturbative strategies to probe the large-$\zeta$, strongly-interacting regime. In such cases, the PDF we can infer with our perturbative method is not an accurate characterization of the actual PDF. It is just evidence that the non-Gaussianity of $\delta\phi$ is non-negligible. Instead, within the naive regime in which perturbation theory is applicable (and thus our method too), milder modifications to the tail of the PDF are possible. Although we have not studied in detail the phenomenological implications in this regime (e.g.\ for the abundance of PBH), it can be relevant for model-building.  

Finally, we remark that the procedure described in this paper can be generalized, as it allows to systematically include non-Gaussianities of $\delta \phi$ in the input for the non-linear relation with $\zeta$ up to an arbitrary order in perturbation theory, provided that the latter holds. 

Our results underline the importance of developing non-perturbative methods to study the tail of the PDF of the curvature fluctuation.

\vspace{0.3cm}

\mysection{Acknowledgments}

{\small We thank Lucas Pinol for discussions and Ashley Wilkins for useful comments. The work of GB, JGE and APR has been funded by the following grants: PID2021-124704NB-I00 funded by MCIN/AEI/10.13039/501100011033 \sloppy and by ERDF A way of making Europe, CNS2022-135613 MICIU/AEI/10.13039/501100011033 and by the European Union NextGenerationEU/PRTR, and Centro de Excelencia Severo Ochoa CEX2020-001007-S funded by MCIN/AEI/10.13039/501100011033.
JGE is supported by a PhD contract {\it contrato predoctoral para formaci\'on de doctores} (PRE2021-100714) associated to the aforementioned Severo Ochoa grant, CEX2020-001007-S-21-3. APR has been supported by Universidad Aut\'onoma de Madrid with a PhD contract {\it contrato predoctoral para formaci\'on de personal investigador (FPI)}, call of 2021, and a grant \textit{Programa de ayudas UAM-Santander para la movilidad de j\'ovenes investigadores}, call of 2024. APR thanks DESY for hospitality during part of the realization of this work. TK, MP and JR acknowledge support by the Deutsche Forschungsgemeinschaft (DFG, German Research Foundation) under Germany's Excellence Strategy – EXC 2121 “Quantum Universe” - 390833306. 
MP acknowledges support by the DFG Emmy Noether Grant No. PI 1933/1-1. 
}

\appendix

\section{The bispectrum}
\label{app:bispectrum}

In this appendix we provide the full expression for the bispectrum, accounting for all terms in the interaction Lagrangian at cubic order, Eq.\ (\ref{eq:fullL}), and we show that the bispectrum in the squeezed limit freezes after horizon crossing. The intrinsic contribution to the bispectrum can be expressed as
\begin{equation}
\mathcal{I}({\bm p},{\bm q},{\bm k})=2\,{\rm Im}\left[\int_{-\infty_-}^t \, \diff t'\,\varphi_p(t)\varphi_q(t)\varphi_k(t)\sum_{n=0}^4 C_n^*(t')\right],
\end{equation}
where $-\infty_- = -\infty(1-i\omega)$ is responsible for projecting onto the interaction vacuum and the coefficients $C_{n}$ can be expressed in terms of the coefficients $c_n$, defined in Eq.~(\ref{eq:smallc}), via
\begin{align}
C_0(t)
&\equiv
6M_P^2\,c_0(t)a(t)^3H(t)^2\varphi_p(t)\varphi_q(t)\varphi_k(t),
\\
C_1(t)
&\equiv
2M_P^2\,c_1(t)a(t)^3\varphi_p(t)\dot{\varphi}_q(t)\dot{\varphi}_k(t)+(p\leftrightarrow q,k),
\\
C_2(t)
&\equiv
M_P^2\,c_2(t)a(t)^3\dot{\varphi}_p(t)\bigg[\frac{{\bm q}\cdot{\bm k}}{k^2}\varphi_q(t)\dot{\varphi}_k(t)+(q\leftrightarrow k)\bigg]+(p\leftrightarrow q,k),
\\
C_3(t)
&\equiv
-2M_P^2\,c_3(t)a(t)({\bm q}\cdot{\bm k})\varphi_p(t)\varphi_q(t)\varphi_k(t)+(p\leftrightarrow q,k),
\\
C_4(t)
&\equiv
2M_P^2\,c_4(t)a(t)^3\frac{({\bm q}\cdot{\bm k})^2}{q^2k^2}\varphi_p(t)\dot{\varphi}_q(t)\dot{\varphi}_k(t)+(p\leftrightarrow q,k),
\end{align}
where the prefactors correspond to the symmetry factors of each term. For the purpose of analyzing the consistency relation of the bispectrum in the squeezed limit \cite{Maldacena:2002vr}, we take $k\ll p$.	 Then
\begin{equation}
\mathcal{I}(k,p)=2\,{\rm Im}\left\{\int_{-\infty_-}^t \, \diff t'\,\Big[
A(t')\varphi_k^*(t')\varphi_p^*(t')^2
+B(t')\varphi_k^*(t')\dot{\varphi}_p^*(t')^2
+C(t')\varphi_p^*(t')\dot{\varphi}_p^*(t')\dot{\varphi}_k^*(t') \Big]
\right\},
\end{equation}
with
\begin{equation}
\frac{A}{M_P^2}\equiv 6a^3H^2c_0+2p^2a\,c_3,
\qquad
\frac{B}{M_P^2}\equiv 2a^3c_1+2a^3c_4,
\qquad
\frac{C}{M_P^2}\equiv 4a^3c_1-2a^3c_2+4\frac{({\bm k}\cdot{\bm p})^2}{k^2p^2}a^3c_4.
\end{equation}

Since $\varphi_k$ is already frozen in the time interval of interest (around the horizon crossing of $p$), one can use $\varphi_k(t')\simeq \varphi_k(t)$ and $\dot{\varphi}_k=0$,  and the integral simplifies as
\begin{equation} \label{eq:Bsqueezed1}
\frac{\mathcal{I}(k,p)}{|\varphi_k(t)|^2|\varphi_p(t)|^2}
=
2\,{\rm Im}\left\{\frac{\varphi_p(t)^2}{|\varphi_p(t)|^2}\int_{-\infty_-}^t \diff t' \Big[
A(t')\varphi_p^*(t')^2
+B(t')\dot{\varphi}_p^*(t')^2 \Big]
\right\}.
\end{equation}
The bispectrum is related to the spectral index of the power spectrum via the consistency relation (\ref{eq:consistency}). Since the power spectrum freezes after the USR phase, the squeezed three-point function $\mathcal{I}(k,p)$ should freeze as well. From Eq.\,(\ref{eq:Bsqueezed1}) it is not obvious that $\mathcal{I} (k,p)$ is time-independent. However, this can be shown analytically in a model-independent way.

One can integrate the term $B(t')\dot{\varphi}_p^*(t')^2$ by parts twice and use the equation of motion of $\varphi$ in addition to the quantization condition ${\rm Im}[\varphi_p(t) \dot{\varphi}_p^*(t) ] = (4 a^3 \epsilon)^{-1}$ to eliminate the dependence on $\dot{\varphi}_p$. This gives
\begin{equation} \label{eq:Bsqueezed2}
	\frac{\mathcal{I} (k,p)}{|\varphi_k(t)|^2|\varphi_p(t)|^2}
	= - \epsilon(t) -2\eta(t) +
	2M_P^2\,{\rm Im} \bigg\{\frac{\varphi_p(t)^2}{|\varphi_p(t)|^2}\int_{-\infty_-}^t a^3 H^2 \bigg[\mathcal{C}(t')-4\epsilon^2\bigg(\frac{p}{aH} \bigg) ^2 \bigg]  \varphi_p^{* 2}(t') \diff t'   \bigg\},
\end{equation}
with
\begin{equation}
	\mathcal{C} \equiv  -2\epsilon \eta \Big[ \epsilon^2 + \epsilon_3\left(3-2\eta + \epsilon_3 + \epsilon_4 \right) - \epsilon\left(3-4\eta+2\epsilon_3 \right)   \Big]  \,.
\end{equation}
Once the modes with momentum $p$ are well outside the horizon, $p\ll a H$, at a time $t'<t$, we obtain analytically and in general that
\begin{equation} 
	{\rm Im} \bigg[\frac{\varphi_p(t)^2 \varphi_p^*(t')^2}{|\varphi_p(t)|^2} \bigg] = 2\, {\rm Im}\left[ \frac{\varphi_p(t)\varphi_p^*(t')}{|\varphi_p(t)|}  \right] {\rm Re}\left[ \frac{\varphi_p(t)\varphi_p^*(t')}{|\varphi_p(t)|}   \right] \, = 2\, {\rm Im}\left[ \varphi_p(t) \varphi_p^*(t') \right] + \mathcal{O}\bigg[ \dfrac{p}{a(t)H(t)} \bigg]. 
\end{equation}
By differentiating twice with respect to time and by using the equation of motion for $\varphi_p$ in the super-horizon limit, we arrive to the result
\begin{equation} \label{eq:Immodessuperh}
	{\rm Im} \bigg[\frac{\varphi_p(t)^2 \varphi_p^*(t')^2}{|\varphi_p(t)|^2} \bigg] = -\dfrac{1}{2M_P^2} \int_{t'}^t \left(\dfrac{1}{a^3\epsilon }\right) {\rm d} t'' + \mathcal{O}\bigg[ \dfrac{p}{a(t)H(t)} \bigg] \,.
\end{equation}
The time integral in Eq.~(\ref{eq:Bsqueezed2}) can be performed by splitting it into two pieces: one up to $t_*$ and one from $t_*$ to $t$ such that the $p$-modes are well outside the horizon for all $t>t_*$. In the second interval one can use the relation (\ref{eq:Immodessuperh}), such that
\begin{align}
\frac{\mathcal{I} (t|k,p)}{|\varphi_k(t)|^2|\varphi_p(t)|^2} = \frac{\mathcal{I} (t_*|k,p)}{|\varphi_k(t_*)|^2|\varphi_p(t_*)|^2} - \epsilon(t) + \epsilon(t_*) -2\big[\eta(t) -\eta(t_*)\big] - \int_{t_*}^t a^3H^2 \mathcal{C}  {\rm d} t' \int_{t'}^t \left( \dfrac{1}{a^3\epsilon } \right) {\rm d}t'' \,,
\end{align}
where we neglected terms suppressed by powers of $p/(aH)$. Noticing that $a^3 \mathcal{C} = \diff \left[-2a^3\epsilon\eta H (\epsilon_3-\epsilon) \right]/\diff t $ and integrating by parts, this expression simplifies to 
\begin{align}
	\frac{\mathcal{I} (t|k,p)}{|\varphi_k(t)|^2|\varphi_p(t)|^2} = \frac{\mathcal{I} (t_* |k,p)}{|\varphi_k(t_*)|^2|\varphi_p(t_*)|^2} + \Big[a^3\epsilon\eta H (\epsilon_3-\epsilon) \Big] \bigg|_{t_*}  \int_{t_*}^t \left( \dfrac{1}{a^3\epsilon } \right) {\rm d} t' \,.
\end{align}

We emphasize that this result is general and that no slow-roll approximation has been made in the derivation. Its interpretation is the following: $\mathcal{I} (t|k,p)$ has a constant and a decaying solution once the modes are outside the horizon. The decaying solution of $\mathcal{I} (t|k,p)$ decays in the same way as the decaying mode of $\varphi_p$ does outside the horizon, as it should. The conditions that guarantee that $\varphi_p(t)$ is constant outside the horizon also ensure that $\mathcal{I} (t|k,p)$ freezes in the same manner.

\section{Failure of the Gram-Charlier series}
\label{app:gram-charlier}

In this appendix we review the Gram-Charlier (or Edgeworth) series and show that it is, in general, not a good approach to calculate the tail of a distribution if the latter deviates significantly from a Gaussian.

The Gram-Charlier expansion \cite{cramer} consists in approximating a non-Gaussian distribution $P_{\rm NG}(x)$ as a series of corrections to a Gaussian distribution $P_{\rm G}(x)$ of variance $\sigma$,
\begin{equation}
	P_{\rm NG}(x)
	=
	P_{\rm G}(x)\bigg[1+\sum_{n=3}^\infty a_nH_n\big(x/\sqrt{2}\sigma\big)\bigg],
	\label{eq:gram_charlier}
\end{equation}
where $H_n$ denotes the $n$-th order Hermite polynomial
\begin{equation}
	H_n(x)=(-1)^ne^{x^2}\frac{\diff^n}{\diff x^n}e^{-x^2}.
\end{equation}
Using the orthogonality of the Hermite polynomials with respect to a Gaussian weight,
\begin{equation}
	\int_{-\infty}^\infty H_m(x)H_n(x)e^{-x^2}\diff x=\delta_{mn}2^nn!\sqrt{\pi}\,,
\end{equation}
one can obtain the coefficients $a_n$ of the Gram-Charlier expansion:
\begin{equation}
	a_n=\frac{1}{2^n n!}\int_{-\infty}^\infty H_n\big(x/\sqrt{2}\sigma\big)P_{\rm NG}(x) \diff x\,.
	\label{eq:gc_coefficients}
\end{equation}

\begin{figure}
	\centering
	\includegraphics[width=0.5\textwidth]{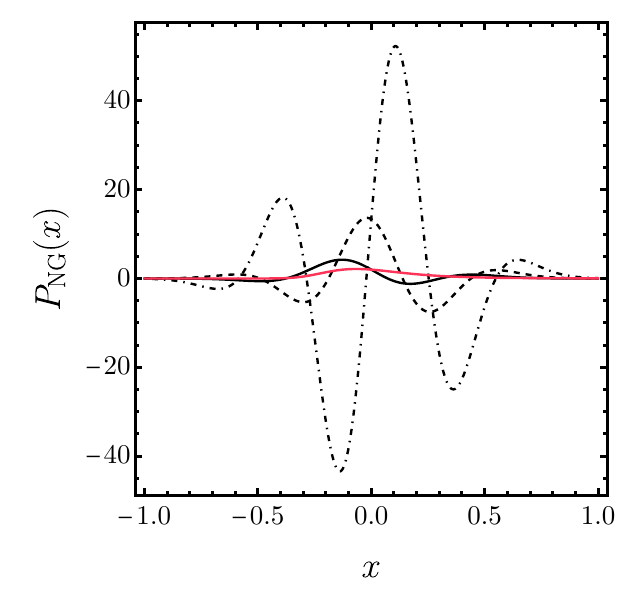}
	\caption{
		\it Non-convergence of the Gram-Charlier series for the PDF in Eq.\,(\ref{eq:pdf_r}). The PDF is shown in red, and the Gram-Charlier series in Eq.\,(\ref{eq:gram_charlier}) up to third, fourth and fifth order is depicted by the solid black, dashed black, and dot-dashed black lines, respectively.
	}
	\label{fig:gram_charlier}
\end{figure}

It is important to remark that this expansion suffers from a series of shortcomings. One is that the PDF resulting from truncating the series at some given order is in general not positive definite. Another is that convergence is in general not guaranteed for an arbitrary PDF. A well-known result \cite{cramer} is that a sufficient condition for the series to converge is that
\begin{equation}
	\int_{-\infty}^\infty e^{x^2/4} P_{\rm NG}(x)\diff x<\infty,
\end{equation}
but this is a very strong condition that is not satisfied for a generic PDF.

As a simple example of a situation in which the above series does not converge, we can use the PDF in Eq.\,(\ref{eq:pdf_r}). We choose $\sigma=0.2$ and $\eta_{\rm CR}=-2$. We find the coefficients via Eq.\,(\ref{eq:gc_coefficients}) and show the resulting PDF truncated order by order in Figure~\ref{fig:gram_charlier}. It is apparent that the convergence of the series becomes progressively worse as more terms are added. Indeed, the Gram-Charlier series is, in general, not an appropriate way to reconstruct a PDF unless the latter is close enough to a Gaussian. {Notice that the PDF from the Gram-Charlier series even becomes locally negative at certain perturbative orders, something that by construction does not happen with the procedure we have put forward in this work.}

\bibliography{mybib2}

\end{document}